\documentclass[a4,english]{smfart}
\usepackage{amsmath,amssymb}%

\usepackage{graphicx}  % standard LaTe\changedX  graphics tool

\usepackage{color}

\usepackage{psfrag}
\usepackage{epsfig}
\usepackage{epstopdf}
\usepackage{wrapfig}
\usepackage{a4}
\usepackage{amssymb,latexsym}%,amsmath}
\usepackage[mathscr]{eucal}
\usepackage{cite}
\usepackage{url}

\makeindex

\DeclareFontFamily{OT1}{rsfs}{} \DeclareFontShape{OT1}{rsfs}{m}{n}{
<-7> rsfs5 <7-10> rsfs7 <10-> rsfs10}{}
\DeclareMathAlphabet{\mycal}{OT1}{rsfs}{m}{n}

\newcommand{\Ten}{{\mbox{T}}}

\newcommand{\HC}{\mathbb{H}^2_{\mathbb{C}}}

\newcommand{\newv}{\zeta}
\newcommand{\neww}{\vartheta}

\begin {document}

\frontmatter

\author [J.L.~Costa]{Jo\~ao~Lopes~Costa}
\address {Lisbon University Institute (ISCTE); Mathematical Institute and Magdalen College, Oxford}
\email {jlca@iscte.pt}
%\urladdr {}

%This mnote stuff should  be on the top of the file
%
%
\newcommand{\mnote}[1]{}%
%{\protect{\stepcounter{mnotecount}}$^{\mbox{\footnotesize $%
%\!\!\!\!\!\!\,\bullet$\themnotecount}}$ \marginpar{%\color{red}
%\raggedright\tiny\em $\!\!\!\!\!\!\,\bullet$\themnotecount: #1} }

\newcommand{\mcEp}{{\mcE^+}}

\newcommand{\kk}[1]{}%{\mnote{{\bf If we consider the KK case:} #1}}

\newcommand{\ohyp}{\,\,\overline{\!\!\hyp}}
\newcommand{\pohyp}{\partial\ohyp}

\newcommand{\regular}{$I^+$--regular}%\newchange{terminology changed (macro, can be restored by resetting to ``regular")}}

\newcommand{\llambda}{\lambda}

\newcommand{\eean}{\nonumber\end{eqnarray}}

%{(\dgt\cup \zh)}
%{\dgt\cup \zh}

\newcommand{\Ndh}{N_{\mbox{\scriptsize\rm  dh}}}

%{S_{0,p}}

\newcommand{\Mtext}{\Sext}%{M_{\mbox{\scriptsize \rm ext}}}

\newcommand{\mcMext}{\Mext}

\newcommand{\mcA}{\mycal A}

\newcommand{\kl}[1]{{(#1)}}

\newcommand{\Uone}{{\mathrm{U(1)}}}

 % The next ones are HR

\newcommand{\jlca}[1]{\mnote{{\bf jlca:} #1}}

\newcommand{\bbR}{\mathbb{R}}

%{{\mycal Doc}}
%
%

\def\scra{{\mycal A}}

\def\e{\wedge}

\def\K0{\phi^{K_0}}

\def\X.{\phi^{X}\cdot}
%{|K_0\e...\e K_{D-3}|^2}

%\newcommand{\Ric}{\operatorname{Ric}}

{\catcode `\@=11 \global\let\AddToReset=\@addtoreset}
\AddToReset{equation}{section}
\renewcommand{\theequation}{\thesection.\arabic{equation}}

\newcommand{\fourg}{{\mathfrak g }}

\newcommand{\nopcite}[1]{}

%{{\widehat \riemg}}
%{{\widehat \riemgz}}

%\newcommand{\<}{\langle}
%\renewcommand{\>}{\rangle}

%\renewcommand{\hbar}{{\overline \riemgz}}

%\newcommand{\letters}
% {\renewcommand{\theenumi}{\alph{enumi}}
%   \renewcommand{\labelenumi}{(\theenumi)}}
%\newcommand{\romanletters}
  {\renewcommand{\theenumi}{\roman{enumi}}
   \renewcommand{\labelenumi}{(\theenumi)}}

\newcommand{\mcE}{{\mycal E}}
\newcommand{\mcC}{{\mycal C}}

\newcommand{\abs}[1]{\left\vert#1\right\vert}

\newcommand{\Lie}{\EuScript L}
\newcommand{\nablash}{\nabla{\kern -.75 em
     \raise 1.5 true pt\hbox{{\bf/}}}\kern +.1 em}
\newcommand{\Deltash}{\Delta{\kern -.69 em
     \raise .2 true pt\hbox{{\bf/}}}\kern +.1 em}
\newcommand{\Rslash}{R{\kern -.60 em
     \raise 1.5 true pt\hbox{{\bf/}}}\kern +.1 em}

\newcommand{\Ric}{\operatorname{Ric}}

%

%\newcommand{\fourg}{{}^{4}g}

 % exterior differential

\newcommand{\hyp}{{\mycal S}}%{\Sigma}{{\mycal S}}

%{\,\,\,\overline{\!\!\!\mycal S}}

 % koneksja tla
 % metryka tla

 %{\pi}} % kontrawariantna gestosc metryki
                            %na czasoprzestrzeni
 % ped sprzezony do kontrawariantnej gestosci
                    % metryki (A duze w innych pracach JK i JJ)
%\newcommand{\E}[1]{{\rm e}^{#1}}
%\newcommand{\kolo}[1]{\stackrel{\circ}{#1}}
%\newcommand{\ve}{\varepsilon}
 %three dimensional metric in
                                %space-time, used for the inverse

\newcommand{\threeg}{\gamma}

 % determinant of the
                                             % three dimensional
                                             % metric pulled back
                               %to the model space
 % the lapse function on the model space
 % the shift vector field on the model
                             % space
 %three dimensional ADM momentum pulled back
                               %to the model space
 % conformally rescaled metric
 % standard round metric on the two
                              % sphere

 % pochodna metryki tla
 % pochodna metryki tla
 %{{}^{n-1}M} %the n-1 dimensional manifold

\newcommand{\mcM}{{\mycal M}}

\newcommand{\bea}{\begin{eqnarray}}
\newcommand{\beaa}{\begin{eqnarray*}}
\newcommand{\bean}{\begin{eqnarray}\nonumber}

\newcommand{\bel}[1]{\begin{equation}\label{#1}}
\newcommand{\beal}[1]{\begin{eqnarray}\label{#1}}
\newcommand{\beadl}[1]{\begin{deqarr}\label{#1}}
\newcommand{\eeadl}[1]{\arrlabel{#1}\end{deqarr}}
\newcommand{\eeal}[1]{\label{#1}\end{eqnarray}}
\newcommand{\eead}[1]{\end{deqarr}}
\newcommand{\eea}{\end{eqnarray}}
\newcommand{\eeaa}{\end{eqnarray*}}

\newcommand{\be}{\begin{equation}}
\newcommand{\ee}{\end{equation}}

\newcommand{\tr}{\mbox{\rm tr}\,}
%{\mbox{\rm \scriptsize ext}\,}

\newcommand{\eq}[1]{\eqref{#1}}
\newtheorem{defi}{\sc Coco\rm}[section]
\newtheorem{theorem}[defi]{\sc Theorem\rm}
\newtheorem{Theorem}[defi]{\sc Theorem\rm}

\newtheorem{Conjecture}[defi]{\sc Conjecture\rm}

\newtheorem{Definition}[defi]{\sc Definition\rm}

\newtheorem{Proposition}[defi]{\sc Proposition\rm}

%\newtheorem{cor}[defi]{\sc Corollary\!\rm}%

%\theoremstyle{remark}
%\theorembodyfont{\upshape}

%\newtheorem{example}[defi]{{\sc Example}\rm}
%\newtheorem{remark}[defi]{{\sc Remark}\rm}

%%%%%%%%%%%%%%%%%%%%%%%%%%%%%%%%%%%%%%%%%%%%%%%%%%%%%

%\newcommand{\proof}{\noindent {\sc Proof:\ }}

\def \C{\mathbb{C}}
\def \R {\Reel}

%\newcommand{\T}{\Bbb{T}}

%\newcommand{\Reel}[0]{\mbox{$\mathbb{R} $}}

%\newcommand{\Hyp}[0]{\mbox{$\mathbb{H} $}}

%\newcommand{\Nat}[0]{\mbox{$\mathbb{N} $}}

%\newcommand{\Sphere}[0]{\mbox{$\mathbb{S} $}}

%\newcommand{\Reel}[0]{\mbox{$\mathit{R} $}}

%\newcommand{\Hyp}[0]{\mbox{$\mathit{H} $}}

%\newcommand{\Nat}[0]{\mbox{$\mathit{N} $}}

%\newcommand{\Sphere}[0]{\mbox{$\mathit{S} $}}

%%%%%%%%%%%%%%%%%%%%%%% notes en marge numerotees %%%%%%%%%%%%%%%%%%%

\newcounter{mnotecount}[section]

\newcommand{\ednote}[1]{}%{\mnote{#1}}

 % The next ones are HR%
%\newcommand{\Sm}{\ensuremath{\Sigma_{-}}}
%\newcommand{\No}{\ensuremath{N_{1}}}
%\newcommand{\Nt}{\ensuremath{N_{2}}}
%\newcommand{\Nth}{\ensuremath{N_{3}}}

%\definecolor{turquoise}{rgb}{0.25,0.8,0.7}
\definecolor{bluem}{rgb}{0,0,0.5}

%Il existe deux repères pour cela :
%+ cyan, magent, yellow, black
\definecolor{mycolor}{cmyk}{0.5,0.1,0.5,0}
\definecolor{michel}{rgb}{0.5,0.9,0.9}
%\newcmykcolor{le_nom_de_la_couleur}{w x y z}
%avec w,x,y,z entre 0.0 et 1.0

%+ red,green, blue et la commande :
\definecolor{turquoise}{rgb}{0.25,0.8,0.7}
\definecolor{bluem}{rgb}{0,0,0.5}

\definecolor{MDB}{rgb}{0,0.08,0.45}
\definecolor{MyDarkBlue}{rgb}{0,0.08,0.45}

\definecolor{MLM}{cmyk}{0.1,0.8,0,0.1}
\definecolor{MyLightMagenta}{cmyk}{0.1,0.8,0,0.1}

\definecolor{HP}{rgb}{1,0.09,0.58}

\newcommand{\opp}[1]{}%{\label{{\mnote{{\color{HP} open problem}}}

\newcommand{\Sext}{\hyp_{\mathrm{ext}}}
\newcommand{\Mext}{\mcM_{\mathrm{ext}}}

\newcommand{\doc}{\langle\langle \mcMext\rangle\rangle}

%{\,\,\,\,\widetilde{\!\!\!\!\cM}}

%\newcommand{\tg}{{\tilde g}}

%{\hskip10pt {\overline{\phantom{I}}\hskip-15pt{\mycal M}}}

%\newcommand{\cL}{{\mycal  L}}

 %background Riemannian metric
 %identity matrix
 %constants

%\def\ba{\begin{eqnarray}}
%\def\ea{\end{eqnarray}}

%\def\half {{1\over 2}}

\def\emph#1{{\it #1}}
\def\textbf#1{{\bf #1}}

\def\tr{\mbox{tr}}

\def\AA{{\mycal  A}}

\def\Lie{{\mycal  L}}

\def\L{{\bf L}}
\def\O{{\bf O}}

\def\R{{\mathbb R}}
\def\U{{\bf U}}

\def\K{{\bf K}}

\newcommand{\changedX}{K}

\def\2{{\overline 2}}

\newcommand{\beqa}{\begin{eqnarray}}
\newcommand{\eeqa}{\end{eqnarray}}

{\catcode `\@=11 \global\let\AddToReset=\@addtoreset}
\AddToReset{figure}{section}
\renewcommand{\thefigure}{\thesection.\arabic{figure}}

%\thanks {}%sponsors

\title{On the classification of stationary electro-vacuum black holes}

\begin {abstract}
We obtain a classification of stationary, $I^+$--regular,
non-degenerate and analytic electro-vacuum space-times in terms of
Weinstein solutions. In particular, for connected horizons, we prove
uniqueness of the Kerr-Newman black holes.
\end {abstract}
%subject classification
 \maketitle

\tableofcontents

\mainmatter

\section{Introduction}

We address the following celebrated and long-standing conjecture:

\begin{Conjecture}
 \label{Cubh} Let $(\mcM,\fourg, F)$ be a stationary, asymptotically flat, electro-vacuum,
four-dimensional regular space-time. Then the domain of outer
communications $\doc$ is either isometric  to the domain of outer
communications of a Kerr-Newman space-time or to the domain of outer
communications of a (standard) Majumdar-Papapetrou space-time.
\end{Conjecture}

Arguments to this effect have been given in the
literature~\cite{CarterlesHouches,Ha1,Mazur} (see
also~\cite{Heusler:book,Weinstein96}), with  the hypotheses needed
not always spelled out, and with some notable technical gaps. The
aim of this work is to give continuation to the project initiated
in~\cite{ChrusCosta}, where the vacuum case was considered, and
obtain a precise classification of such electro-vacuum solutions
in the class of analytic space-times
with non-degenerate event horizons, providing detailed filling of
the gaps alluded to above.

%The (pure) vacuum Black Hole Uniqueness Conjecture aims for a
%classification in terms of a single family of solutions, the Kerr
%family. For electro-vacuum we expect a complete classification in
%terms of two distinct families: the Kerr-Newman space-times, a
%charged generalization of the Kerr metrics, are expected to be
%unique within the connected event horizon class (i.e., single black
%hole solutions); the Majumdar-Papapetrou solutions are expected to
%include all {\em regular} solutions with non-connected event horizon
%(many black hole solutions). Although we expect the results
%presented to come a long way in overcoming the difficulties created
%by the presence of the ``technically awkward"\cite{}\jlca{quote
%Carter} degenerate horizons, the main result of the paper concerns
%space-times with all horizons non-degenerate; this eliminates the
%extreme Kerr-Newman solutions as well as the Majumdar-Papapetrou
%family from our classification.

As usual, in mathematical Relativity, part of the challenge posed by
a conjecture is to obtain a precise formulation. In the
case of the ``no-hair" conjectures this non-linearity lies in the
notion of {\em regularity} and it is our opinion that the
non-existence of a precise formalization for this concept has led to
the enclosure of ``hairy" assumptions and technical difficulties
which has made the  state of the art concerning this problem
difficult to assess.\footnote{An illustrative example is given by
the product structure~\eqref{doc minus axis} that, although clear
for Minkowski with the usual $\R\times U(1)$ action by isometries,
seems far from obvious in the generality required. Other examples
are the regularity of the horizon and the asymptotic behavior of the
relevant harmonic maps.} So, we start exactly by collecting our
technical assumptions in the following  (we refer to
Section~\ref{Sprelim} for the necessary intermediary definitions):

\begin{Definition}
 \label{Dmain}
Let $(\mcM,\fourg)$ be a space-time containing an asymptotically
flat end $\Sext$, and let  $\changedX $ be a stationary Killing
vector field  on $\mcM$.
We will say that $(\mcM,\fourg,\changedX)$ is $\mbox{\rm {\regular}}$%
\index{$\mbox{\rm {\regular}}$}
if $\changedX $ is complete, if the domain of outer communications
$\doc$ is globally hyperbolic, and if $\doc$ contains a spacelike,
connected, acausal hypersurface $\hyp\supset\Sext $,%
\index{$\hyp$}
the  closure $\ohyp $ of which is a topological manifold with boundary,
consisting of  the union of a compact set and of a finite number of
asymptotic ends, such that the boundary $ \pohyp:= \ohyp \setminus \hyp$
is a  topological manifold satisfying
\bel{subs}
\pohyp \subset \mcE^+:= \partial \doc \cap I^+(\Mext)
 \;,
 \ee
with $\pohyp$
meeting every generator of $\mcE^+$ precisely once. (See Figure~\ref{fregu}.)
\end{Definition}

Needless to say, all these conditions are satisfied by the
Kerr-Newman and the Majumdar-Papapetrou solutions and, in
particular, by Minkowski and Reissner-Nordstr\"om. For a detailed
discussion of the previous definition and an alternative formulation
of Conjecture~\ref{Cubh} with {\em regular} replaced by a specific
set of weaker conditions see~\cite{ChrusCosta}.

\begin{figure}[ht]
\begin{center} { \psfrag{Mext}{$\phantom{x,}\Mext$}
\psfrag{H}{ } \psfrag{B}{ }
\psfrag{H}{ }
 \psfrag{pSigma}{$\!\!\pohyp\qquad\phantom{xxxxxx}$}
\psfrag{Sigma}{ $\hyp$ }
 \psfrag{toto}{$\!\!\!\!\!\!\!\!\!\!\doc$}
 \psfrag{S}{}
\psfrag{H'}{ } \psfrag{W}{$\mathcal{W}$}
\psfrag{scriplus} {} %{ $\mathcal{I}^+$}
\psfrag{scriminus} {} %{ $\mathcal{I}^-$}
 \psfrag{i0}{}%{ $i^0$}
\psfrag{i-}{ } \psfrag{i+}{}
 \psfrag{E+}{ $\phantom{.}{\mycal E}^+$}
%\resizebox{3in}{!}
{\includegraphics{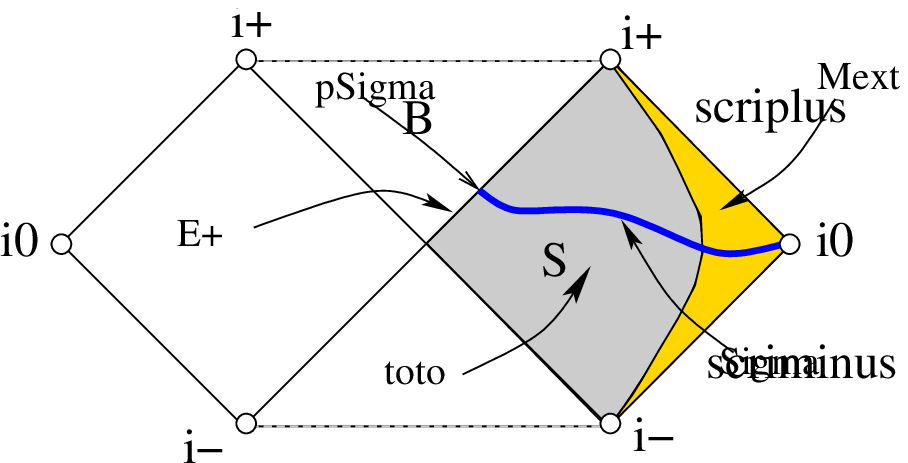}}
}
\caption{The hypersurface $\hyp$ from the definition of \regular ity.
\protect\label{fregu}}
\end{center}
\end{figure}

%%
%\begin{figure}[t]
%\begin{center} {
%\resizebox{3.5in}{!}{\includegraphics{simpleQ.eps}
%\includegraphics{becomes.eps}\includegraphics{doubledQ.eps}}
%}
%\caption{The quotient space and its double.
%\label{Fquot}}
%\end{center}
%\end{figure}
%%

%\jlca{figure missing. There is a conflict between the show label
%package and the way the picture is included.}

%
In this work we establish the following special case of Conjecture~\ref{Cubh}:

\begin{theorem}
 \label{Tubh} Let $(\mcM,\fourg, F)$ be a stationary,
asymptotically flat, {\regular}, electro-vacuum, four-dimensional
analytic space-time, satisfying~\eqref{A decay} and~\eqref{Finv}. If
each component of the event horizon is mean non-degenerate,\jlca{no
definition trough the entire paper} then $\doc$ is isometric to the
domain of outer communications of one of the Weinstein solutions of
Section~\ref{sWeinstein}.  In particular, if the event horizon is
connected and mean non-degenerate, then $\doc$ is isometric  to the
domain of outer communications of a Kerr-Newman space-time.
\end{theorem}

 It should
be emphasized that the hypotheses of analyticity and non-degeneracy
are highly unsatisfactory,  and one believes that they are not
needed for the conclusion. Note that by not allowing the existence
of the ``technically awkward"\cite{Carter99} degenerate horizons we
eliminate extreme Kerr-Newman as well as the Majumdar-Papapetrou
solutions from our classification. One also believes, in accordance
with the statement of Conjecture~\ref{Cubh}, that all solutions with
non-connected event horizon are in the Majumdar-Papapetrou family;
consequently one expects all other (non-connected) Weinstein
solutions,  and in particular the ones referred to in the previous
result, to be singular. We postpone further discussion of this
issues to the final section.

A critical remark comparing our work with the existing literature is
in order; we focus on those points that do not generalize
immediately when passing from pure to electro-vacuum. First of all,
the famous reduction of the Einstein-Maxwell source free equations
to a singular harmonic map problem requires the use of Weyl
coordinates. The local existence of such coordinates has been well
known for some time now, but global existence has, to our knowledge,
either been  part of the ansatz, usually implicitly, or based on
incorrect or incomplete analysis. The main reasons for this
unsatisfactory situation resides in the existing proofs of
non-negativity  of the {\em area function}~\eqref{area} in $\doc$,
and existence of a global cross-section for the $\R\times U(1)$
action again in $\doc$. In~\cite{ChrusCosta} we solved this problems
 for vacuum and in Section~\ref{sWeyl} we present the
necessary adjustments  to extend this global existence result to the
electro-vacuum scenario. Also, no previous work  known to us
establishes the asymptotic behavior, as needed for the proof of
uniqueness, of the relevant harmonic maps. More specifically: the
necessity to control the behavior at points where the horizon meets
the rotation axis, prior to~\cite{ChrusCosta}, seems to have been
neglected; at infinity, which requires special attention in the
electro-vacuum case, part of the necessary estimates were imposed as
extra conditions, beyond asymptotic flatness;~\footnote{See, for
example,  Theorem 2 in~\cite{Mazur08}.} also, an apparent disregard
for the  singular character, at the axis, of the hyperbolic
distance~\eqref{dist} between the maps, even at large distance,
appears to be the norm.  A detailed asymptotic analysis is carried
out in Section~\ref{sBoundary}.

We also note that a considerable part of the foundations of the
theory underlying the desired classification depend exclusively on
stationarity, $I^+$--regularity and the null energy condition.
Again, this work was carried out in~\cite{ChrusCosta}, where various
results were established under conditions weaker than previously
cited, or were generalized to higher dimensions;  this is of
potential interest for further work on the subject.

\section{Preliminaries}
\label{Sprelim}

An {\em electro-vacuum} space-time is a triple $(\mcM, \fourg, F)$,
assembled by a $(n+1)$--dimensional Lorentzian manifold $(\mcM,
\fourg)$ endowed with a 2-form $F$, that  satisfies the {\em source
free Einstein-Maxwell} field equations
\bel{fieldeq1}
 \text{Ric}-\frac{1}{2}\text{R}\fourg=2\,\text{T}_F\;,
\ee
\begin{equation}
\label{fieldeq2}
 F=dA\;,
\end{equation}
\begin{equation}
\label{fieldeq3} d*F=0\;,
\end{equation}
where Ric is the Ricci curvature tensor of the metric  $\fourg$, R
its scalar curvature and T$_F$ is the energy-momentum tensor of the
{\em electromagnetic} 2-form $F$,
\begin{align}
\label{setensor}
 \text{T}_F(u,v) := \fourg(i_uF,  i_vF)-\frac{1}{2}|F|^2\fourg(u,v)\;.
\end{align}

\bigskip

A space-time will  be said to possess an \emph{asymptotically flat
end } if $\mcM$ contains a spacelike hypersurface $\Mtext$
diffeomorphic to $\R^n\setminus B(R)$, where $B(R)$ is an open
coordinate ball of radius $R$, with the following properties: there
exists a constant $\alpha>0$ such that, in local coordinates on
$\Mtext$ obtained from $\R^n\setminus B(R)$, the metric $\threeg$
induced by $\fourg$ on $\Mtext$, and the extrinsic curvature tensor
$K_{ij}$ of $\Mtext$, satisfy the fall-off conditions
\beal{falloff1}
 & \threeg_{ij}-\delta_{ij}=O_k(r^{-\alpha})\;,  \qquad
  K_{ij}=O_{k-1}(r^{-1-\alpha})\;,
 \eeal{falloff2}
for some $k \geq 1$, where we write $f=O_k(r^{\alpha})$ if $f$
satisfies
\bel{okdef}
  \partial_{k_1}\ldots\partial_{k_\ell}
f=O(r^{\alpha-\ell})\;, \quad 0\le \ell \le k
 \;.
\ee
In connection with the field equations we also request the following
decay rate for the electromagnetic potential
 \bel{A
decay} \quad A_{\mu}=O_k(r^{-{\alpha}})\;.\ee

A Killing vector $K$ is said to be {\em complete} if for every $p\in
\mcM$ the orbit $\phi_t[K ](p)$ of $\changedX $ is defined for all
$t\in \R$, i.e., if (the flow of) $K$ generates an action of $\R$ by
isometries; in an asymptotically flat context, $\changedX $ is
called \emph{stationary} if it is timelike at large distances. The
exterior region $\Mext$
\index{$\Mext$}
and the \emph{domain of outer communications}%
\index{domain of outer communications}
$\doc$ are  then defined as%
%\footnote{Recall that $I^-(\Omega)$, respectively $J^-(\Omega)$, is
%the set covered by past-directed timelike, respectively causal,
%curves originating from $\Omega$, while $\dot I^- $ denotes the
%boundary of $I^-$, etc. The sets $I^+$, etc., are defined as $I^-$,
%etc., after changing time-orientation.}
%%
%(compare Figure~\ref{FPast})
%%
%\begin{figure}[t]
%\begin{center} { \psfrag{Mext}{\Large$\,\Mext$}
%\psfrag{H}{ } \psfrag{B}{ } \psfrag{pasthorizon}{\Large
%$\!\!\!\!\!{\partial I^+(\Mext)}$ }
% \psfrag{pSigma}{$\!\!\pohyp\qquad\phantom{xxxxxx}$}
%\psfrag{Sigma}{\Large $\!\Sext$}
% \psfrag{toto}{\Large$\!\!\!\!\!\!\!\!\!\!\!\!\!\!\!\!\! I^-(\Mext)$}
% \psfrag{S}{}
% \psfrag{future}{\Large $\!\!\!\!\!{I^+(\Mext)}$}
%\psfrag{H'}{ } \psfrag{W}{$\mathcal{W}$}
%\psfrag{scriplus} {} %{ $\mathcal{I}^+$}
%\psfrag{scriminus} {} %{ $\mathcal{I}^-$}
% \psfrag{i0}{}%{ $i^0$}
%\psfrag{i-}{ } \psfrag{i+}{}
% \psfrag{E+}{\Large $\!\!\!\!\!{\partial I^-(\Mext)}$}
%\resizebox{2.3in}{!}{\includegraphics{PastScriPlus.eps}}
%\resizebox{2.3in}{!}{\includegraphics{FutureScriMinus.eps}} }
%\caption{$\Sext$, $\Mext$, together with the future and  the past of
%$\Mext$. One has $\Mext\subset I^\pm(\Mext)$, even though this is
%not immediately apparent from the figure. The domain of outer
%communications is the intersection $ I^+(\Mext)\cap I^-(\Mext)$,
%compare Figure~\ref{fregu}. \label{FPast}}
%\end{center}
%\end{figure}
%%
%%
\bel{docdef}
 \doc = I^+(\underbrace{\cup_t \phi_t(
 \Sext )}_{=:\Mext})\cap  I^-(\cup_t \phi_t
 (\Sext ))
 \;,
\ee
with the {\em event horizon} being
\bel{event horizon}
 \mcE:=\partial\doc
\;\;\;;\;\;\;\mcE^{\pm}:=I^{\pm}(\Mext)\cap\mcE \;.\ee

 One expects stationary electro-vacuum space-times
to be \emph{static} or {\em stationary-axisymmetric}: {\em static}
meaning that the stationary Killing vector is
hypersurface-orthogonal, i.e.,
\bel{static} dK^{\flat}\wedge K^{\flat}=0\;, \ee
where $K^{\flat}=\fourg(K,\cdot)$, and {\em stationary-axisymmetric}
corresponding to the existence of a second complete Killing vector
$K_\kl {1}$, which together with the stationary Killing vector
$K_\kl 0:=K$ generate an $\R\times\U(1)$ action by isometries. In
connection with the field equations we require the
electromagnetic field to be invariant under the flow of the relevant
Killing vectors
\bel{Finv} \Lie_{K_\kl {\mu}}F=0 \;.\ee

In the stationary and asymptotic flat scenario one is able to choose
adapted coordinates so that the metric can, in a neighborhood of
infinity, be written as
\beal{gme1}
 &\fourg  =
 -V^2(dt+\underbrace{\theta_i dx^i}_{=\theta})^2 +
 \underbrace{\threeg_{ij}dx^i dx^j}_{=\threeg}\;,
\eeal{gme2}
with
\bel{foff0}
 K=\partial_t\Longrightarrow\partial_t V = \partial_t \theta_i = \partial_t \threeg_{ij}=0
\;; \ee
since we are also assuming electro-vacuum we get the following
improvement of the original decay rates~\cite[Section
1.3]{ChruscielNo},\jlca{This gives $\alpha$ but not $k$ explicitly.
But is what Piotr says in~\cite{Chstaticelvac} pg692. Also $k\geq
1$}
\bel{foff}
 \threeg_{ij}-\delta_{ij}=O_{\infty}(r^{-1})\;, \quad
  \theta_{i } =O_{\infty}(r^{-1})\;, \quad V-1= O_{\infty}(r^{-1})
 \;,
\ee
and
 \bel{foff A} \quad A_{\mu}=O_{\infty}(r^{-1})\;,\ee
 where the infinity symbol means that~\eqref{okdef} holds for
arbitrary $k$.

\section{Weyl coordinates}
\label{sWeyl}

%After doubling across the boundary, one
%obtains an asymptotically flat metric on $\R^2$.
%By~\cite[Proposition~2.3]{ChUone}, for $k\ge 5$ in
%\eq{foff2} there exist global isothermal coordinates for
%$q$:
%%
%\bel{hat coord}q=e^{2\hat u}(d\hat x^2+d\hat z^2)\;, \quad \mbox{with} \
%\hat u\longrightarrow_{\sqrt{\hat x^2+\hat z^2} \to \infty}0\;.\ee
%%
%In fact, $u=o_{k-4}(\hat r^{-1})$. The existence of such coordinates also follows from the uniformization theorem~(see, e.g.,~\cite{Abikoff}),
%but this theorem does not seem to provide the information about the asymptotic behavior in various regimes, needed here, in any obvious way.
%As explained in the proof of~\cite[Theorem~2.7]{ChUone}, the coordinates $(\hat x,\hat z)$ can be chosen so that the rotation axis corresponds
%to $\hat x=0$, with orbit space given by $\mcO=\{\hat x\ge 0\}$.
%%
%\jlca{This is copy paste. It needs to be revised completely}

On a region charted by Weyl coordinates  the source free Einstein-Maxwell equations simplify considerably.
It has been for long expected and recently showed in~\cite{ChrusCosta} that such global chart is available away from the axis
of a stationary
and axisymmetric vacuum domain of outer communications.
In fact the role of the vacuum field equations in the referred analysis --
they imply the {\em orthogonal integrability conditions}~\eqref{intcond} and  allow us to show that,
whenever defined,  the squared root of the {\em area function}~\eqref{area}
is harmonic with respect to the orbit space metric --
is fulfilled by the electro-vacuum field equations.

The first of these well known results, which neither requires $K_\kl
0$ to be stationary, nor  $K_\kl 1$ to be a generator of
axisymmetry,  generalizes  to higher dimensions as follows
(compare~\cite{CarterJMP}):

\begin{Proposition}
\label{xyort} Let $(\mcM,\fourg, F)$ be an $(n+1)$--dimensional
electro-vacuum space-time, possibly with a cosmological constant,
with $n-1$
 commuting Killing vector fields satisfying
$$\Lie_{K_{\kl \mu}} F=0
\;\;\;,\;\mu=0,\ldots,n-2\;.$$
If $n-2$ of the zero sets $\mcA_{\mu}:=\{p\in\mcM \ | \ K_\kl {\mu}|_p=0\}$ are non-empty
 then
 \footnote{By an abuse of notation, we use the same symbols for
 vector fields and for the
associated 1-forms.}
\begin{equation}
\label{intcond} dK_\kl{\mu }\wedge K_\kl{ 0} \wedge \ldots
\wedge
K_\kl{{n-2}} =0 \;\;\;,\;\forall\mu=0,\ldots,n-2\;.
\end{equation}
\end{Proposition}

\begin{proof}
\newcommand{\ourstar}{*}%
 To fix conventions,  we use a Hodge star defined through the
formula
$$
\alpha\wedge \beta = \pm \langle \ourstar  \alpha, \beta\rangle
\mathrm{Vol}\;,
$$
where the plus sign is taken in the Riemannian case, minus  in
our
Lorentzian one, while  $\mathrm{Vol}$ is the volume form. The
following
(well {known}) identities are useful~\cite{Heusler:book};
\begin{equation}
\label{hodge1}
 **\theta=(-1)^{s(n+1-s)-1}\theta
  \;,
  \qquad \forall \theta\in\Lambda^s\;,
\end{equation}
\begin{equation}
\label{hodge2}
 i_X *\theta=*(\theta\e X )\;,\qquad \forall
 \theta\in\Lambda^s\;, \quad X \in\Lambda^1\;.
\end{equation}
Further, for any Killing vector $K$,
\begin{equation}
\label{hodge3}
 [\Lie_K,*]=0\;.
\end{equation}
The Leibniz rule for the divergence $\delta:=*d*$ reads, for
$\theta \in
\Lambda^s$,
 \begin{align*}
\delta(\theta\e K) &=*d*(\theta\e K){\buildrel \eqref{hodge2}
                                      \over
                                      =}*d(i_K*\theta)=*(\Lie_K*\theta-i_Kd*\theta)
                                      \\
          &{\buildrel \eqref{hodge1},\eqref{hodge3}
                                      \over =}**\Lie_K\theta
                                      -*i_K(-1)^{(n+1-s+1)(n+1-(n+1-s+1))-1}**d*\theta
                                      \\
          &=(-1)^{s(n+1-s)-1}\Lie_K\theta-(-1)^{s(n+1-s)-n+1}**(\delta
         \theta\e K)\\
          &=(-1)^{s(n+1-s)-1}\Lie_K\theta+(-1)^{n+1}\delta
          \theta\e K\;.
\end{align*}
Applying this to $\theta=dK$ one obtains
 \begin{align*}
      *d*(dK\e K) &=- \Lie_K dK+(-1)^{n+1}\delta dK\e K     \\
          &= (-1)^{n+1}\delta dK\e K
          \;.
\end{align*}
As any Killing vector is divergence free, we see that
$$
 \delta d K=(-1)^{n}\Delta K=(-1)^n 2\,\tr \nabla^2 K=(-1)^{n+1}2\,i_K \Ric
 \;,
$$
 where $\Delta$ is the Laplace-Beltrami operator. The assumed field equations (with cosmological constant $\Lambda$) imply
$$\Ric=
2\,\Ten_{F}+\frac{2}{n-1}\Lambda \fourg\;,$$
from which
\begin{align*}
*d*(dK\e K) & =(-1)^{n+1}(-1)^{n+1}2\,i_K(2\,\Ten_F+\frac{2}{n-1}\Lambda \fourg)\e K\
\\
&= 2\,\left(2\,i_K\Ten_F\e K +\frac{2}{n-1}\Lambda\, K\right)\e K =
4\,i_K\Ten_F\e K\;.
\end{align*}

Letting $\alpha:=i_{K}F$ for any vector field $X$ we have
 \begin{align*}
\alpha\cdot i_X F &=-*(\alpha\e *i_XF)=-(-1)^{n}*(*i_XF\e \alpha)                \\
          &= (-1)^{n+1}i_{\alpha}**F=(-1)^{n+1}(-1)^{n+1}i_{\alpha}i_X F \\
          &= -F(\alpha,X)
          \;,
\end{align*}
which inserted into~\eqref{setensor}
gives
$$i_KT_F=-i_{\alpha\,}F-\frac{1}{2}|F|^2K\;,$$
and consequently
$$*d*(dK\wedge K)=-4(i_{\alpha}F+\frac{1}{2}|F|^2K)\e K=-4\, i_{\alpha}F\e K=4\,K\e i_{\alpha}F\;.$$

Meanwhile, since (modulo sign)
\begin{align*}
i_{\alpha}K &=\pm*(K\e*\alpha)=\pm*(K\e*i_KF)=\pm*(K\e*i_K**F)
\\
&=
\pm*(K\e**(*F\e K))=\pm*(K\e*F\e K)=0
\;,
\end{align*}
for $\beta:=i_K*F\in \Lambda^{n-2}$, we have
 \begin{align*}
*(\alpha\e\beta) &=(-1)^{1\times(n-2)}*(\beta\e \alpha)=(-1)^{n-2}i_{\alpha}*\beta
    \\
    &=(-1)^{n-2}i_{\alpha}*i_K*F=(-1)^{n-2}i_{\alpha}**(F\e K)
    \\
    &=(-1)^{n-2}i_{\alpha}(-1)^{3(n+1-3)-1}F\e K
    \\
    &=-(i_{\alpha}F\e K+(-1)^2F\e i_{\alpha}K)
          =-(i_{\alpha}F\e K+0)\\
          &=K\e i_{\alpha}F
\end{align*}
which leads to the significant
\begin{equation}
\label{domega} d*(dK\wedge K)= 4\,\alpha\e \beta = 4\,i_KF\e
i_K*F\;.
\end{equation}
 Now, for any two  commuting {Killing} vectors and an arbitrary differential form we have
\begin{align*}
    [\Lie_{K_{\kl \mu}},i_{K_{\kl \nu}}]\theta &=\Lie_{K_{\kl \mu}}(i_{K_{\kl \nu}}\theta)-i_{K_{\kl \nu}}(\Lie_{K_{\kl \mu}}\theta)
          \\
          &=\Lie_{K_{\kl \mu}}[\theta(K_{\kl \nu}\,,\ldots)]-(\Lie_{K_{\kl \mu}}\theta)(K_{\kl \nu}\,,\ldots)
          \\
          &=(\Lie_{K_{\kl \mu}}\theta)(K_{\kl \nu}\,,\ldots)+\theta(\Lie_{K_{\kl \mu}}K_{\kl \nu}\,,\ldots)
          -(\Lie_{K_{\kl \mu}}\theta)(K_{\kl \nu}\,,\ldots)=0
          \;,
\end{align*}
giving us the commutation relation
\begin{equation}
\label{Li-iL}[K_{\kl \mu},K_{\kl \nu}]=0\Longrightarrow [\Lie_{K_{\kl \mu}},i_{K_{\kl \nu}}]=0
\;,
\end{equation}
from which it follows that
 \begin{align*}
  dF(K_{\kl \nu},K_{\kl \mu}) &=di_{K_{\kl \mu}}\alpha_{\kl \nu}=-i_{K_{\kl \mu}}d\alpha_{\kl \nu}+\Lie_{K_{\kl \mu}}\alpha_{\kl \mu}
          \\
          &=-i_{K_{\kl \mu}}(-i_{K_{\kl \nu}}dF+\Lie_{K_{\kl \nu}}F)+i_{K_{\kl \mu}}\Lie_{K_{\kl \mu}}F=0\;,
\end{align*}
where we used the fact that $F$ is exact and invariant under the flow of this Killing vectors.
By the hypothesis on the zero sets, for any pair ${\mu}\neq { \nu}$, we may take $\scra_{\kl \mu}\neq\emptyset$. We then have $F(K_{\kl \mu},K_{\kl \nu})|_{\scra_{\kl \mu}}\equiv0$
and consequently
\begin{equation}
\label{Fij=0} F(K_{\kl \mu},K_{\kl \nu})\equiv 0\,\,,\,\,\forall\,
\mu,{\nu}\in\{0,...,n-2\}
\;.
\end{equation}

A similar computation leads to
\begin{equation}
\label{dualFij=0} i_{K_{\kl \mu}}i_{K_{\kl \nu}}*F=0\,\,,\,\,\forall\,
{ \mu},{ \nu}\in\{0,...,n-2\}
\;.
\end{equation}
Now, let $\omega_\kl \mu$ be the $\mu$'th twist form,
$$\omega_\kl \mu :=
\ourstar (dK_\kl \mu\wedge K_\kl \mu)
 \;.
$$
%;  t
The identity
\begin{align*}
          \Lie_{K_\kl \mu}\omega_\kl \nu&=\Lie_{K_\kl
          \mu}*(dK_\kl \mu\e K_\kl \nu)                \\
          &=*(\Lie_{K_\kl \mu}dK_\kl \nu+dK_\kl
          \nu\e\Lie_{K_\kl \mu}K_\kl \nu)=0\;,
\end{align*}
together with
$$
 \Lie_{K_{\kl {\mu_1}}}(i_{K_{\kl {\mu_2}}}\ldots  i_{K_{\kl
 {\mu_{\ell}}}}\omega_{\kl {\mu_{\ell+1}}})
 =i_{K_{\kl {\mu_2}}}\ldots i_{K_{\kl {\mu_{n-1}}}}\Lie
 _{K_{\kl {\mu_{\ell}}}}\omega_{\kl {\mu_{\ell+1}}}
 =0
 \;,
$$
and Cartan's formula for the Lie derivative, gives
\begin{equation}
\label{dii-iid}
d(i_{K_{\kl {\mu_1}}}\ldots i_{K_{\kl {\mu_{\ell}}}}\omega_{\kl
{\mu_{\ell+1}}})
 =
 (-1)^\ell i_{K_{\kl {\mu_1}}}\ldots i_{K_{\kl
 {\mu_{n-1}}}}d\omega_{\kl {\mu_{\ell+1}}}
 \; .
\end{equation}
We thus have
\begin{align*}
        d*(dK_{\kl {\mu_{0}}}\e K_{\kl {\mu_{0}}}\e\ldots \e
        K_{\kl {\mu_{{n-2}}}})
        &=
        d(i_{K_{\kl {\mu_{ {n-2}}}}}\ldots i_{K_{\kl
        {\mu_{1}}}}*(dK_{\kl {\mu_{0}}}\e K_{\kl {\mu_{0}}}))
        \\
        &=
        (-1)^{n-2}i_{K_{\kl {\mu_{n-2}}}}\ldots i_{K_{\kl
        {\mu_1}}}d\omega_{\kl {\mu_0}}
        \\
        &{\buildrel \eqref{domega}
        \over =}
        (-1)^{n} i_{K_{\kl {\mu_{n-2}}}}\ldots i_{K_{\kl
        {\mu_1}}}4\,\alpha_{\kl {\mu_0}}\e \beta_{\kl {\mu_0}}
        \\
        &=
        4\,(-1)^{n}i_{K_{\kl {\mu_{n-2}}}}\ldots i_{K_{\kl
        {\mu_2}}}
        \\
        & \;\;\;\;\; \;\;\;\;\; \;\;\;\;\; \;\;\;\;\;
        (i_{K_{\kl {\mu_1}}}\alpha_{\kl {\mu_0}}\e\beta_{\kl {\mu_0}}-\alpha_{\kl {\mu_0}}\e i_{K_{\kl {\mu_1}}}\beta_{\kl {\mu_0}})
        \\
        &=
        4\,(-1)^{n} i_{K_{\kl {\mu_{n-2}}}}\ldots i_{K_{\kl
        {\mu_2}}}
        \\
        & \;\;\;\;\; \;\;\;\;\; \;\;\;\;\; \;\;\;\;\;
         (F(K_{\kl {\mu_0}},K_{\kl {\mu_1}})\beta_{\kl {\mu_0}}-\alpha_{\kl {\mu_0}}\e i_{K_{\kl {\mu_1}}}i_{K_{\kl {\mu_0}}}*F)
        \\
        & {\buildrel (\ref{Fij=0},\ref{dualFij=0})
        \over =}0
        \;.
\end{align*}
So the function   $*(dK_{\kl {\mu_0}}\e K_{\kl {\mu_0}}\e
K_{\kl
{\mu_1}}\e\ldots \e K_{\kl {\mu_{n-2}}})$ is constant, and, as before, the
result follows from the hypothesis on the zero sets.

\end{proof}

Noting that a globally hyperbolic, stationary and
asymptotically flat domain of outer communications satisfying the
null energy condition is necessarily simply-connected~\cite{ChWald,
galloway-topology, Galloway:fitopology}, in view of the previous result Theorem~5.6
of~\cite{ChrusCosta} translates to the electro-vacuum setting as:

\begin{Theorem}
 \label{Tdoc}
Let $(\mcM,\fourg, F)$ be a four-dimensional, \regular,
asymptotically flat, electro-vacuum space-time with stationary
Killing vector $K_\kl 0$ and periodic Killing vector $K_\kl 1$,
jointly generating an $\R\times \Uone$ subgroup of the isometry
group of $(\mcM,\fourg)$. If $\doc$ is globally hyperbolic, then the
area function
\bel{area}
 W:= -\det \Big(\fourg(K_\kl \mu, K_\kl
 \nu)\Big)_{\mu,\nu=0,1}
 \;,
\ee
 is non-negative on $\doc$,
vanishing precisely on the union of its boundary with the
(non-empty) set $\{\fourg(K_\kl 1,K_\kl 1)=0\}$.
\end{Theorem}

Away from points where $K_\kl 0\wedge K_\kl 1$ vanishes, which
according to~\cite[Corollary 3.8]{ChrusCosta} correspond, in a
chronological\footnote{No closed timelike curves allowed.} $\doc$,
exactly to axis points
\bel{axis} \mcA:=\{q\in\mcM \ | \ K_\kl 1|_q=0\} \;, \ee
there is a well defined and differentiable local cross-section for
the $\R\times U(1)$ action. We can endow this cross-section with the
orbit space metric
\bel{OSmetst}
 q(Z_1,Z_2)=\fourg(Z_1,Z_2)- h^{\mu\nu}{\fourg(Z_1,K _{\kl \mu})\fourg(Z_2,K _{\kl \nu})}
 \;,
\ee
whenever $h_{\mu\nu}:=\fourg(K_\kl \mu, K_\kl \nu)$ is non-singular.
The established orthogonality conditions allow us to identify,
at least locally, the previous orbit space structure  with a
2-surface orthogonal to the Killing vectors, provided by~\eqref{intcond}, endowed with the induced
metric.
 From this and Theorem~\eqref{Tdoc} we see that $q$ is well
defined and Riemannian throughout $\doc\setminus\mcA$; it is then
well known~\cite{Weinstein96} that
 \bel{rho harmonic}
\Delta_q \sqrt{W}=0\;, \ee
whenever $W$ is non-negative and $q$ is Riemannian, which again is
the case within $\doc\setminus\mcA$.

According to the  Structure Theorem~\cite{ChrusCosta},
$I^+$--regularity allows for the decomposition
\bel{doc decomp} \overline{\doc}\cap
I^{+}(\Mext)=\R\times\overline{\hyp}\;,\ee
with $K_\kl 1$ tangent to $\overline{\hyp}$, a simply-connected
spacelike hypersurface with boundary which is an asymptotically flat
global cross-section for the action generated by the stationary
vector. We are now allowed to use the classification of circle
actions on simply-connected 3-manifolds of Orlik and
Raymond~\cite{Orlik, Raymond} to obtain a global cross-section for
the $\R\times\U(1)$ action in $\doc\setminus\mcA$. Then,
by~\eqref{rho harmonic} and relying on the results
of~\cite{ChUone}, while disallowing the existence of
degenerate\jlca{no definition or reference to!} horizons, we are
able to undertake an analysis leading to
\bel{doc minus axis}\doc\setminus \mcA
  \approx \bbR\times S^1 \times \bbR^+\times \bbR  \;,
  \ee
while showing that this diffeomorphism defines a global coordinate system  $(t,\varphi,\rho,z)$ with
\bel{XYrho}
 K_\kl 0=\partial_t\;,\quad  K_\kl 1=\partial_{\varphi}\;\; \text{ and
}\;\;\rho=\sqrt{W}\;. \ee
After invoking~\eqref{intcond} once more, the
desired global expression for the space-time metric in terms of Weyl
coordinates
\begin{equation}
\label{space-time metric}
\fourg=-\rho^2e^{2\llambda}dt^2+e^{-2\llambda}(d\varphi-w
dt)^2+e^{2u}(d\rho^2+dz^2)\;,
\end{equation}
follows, with
\bel{decay hat u} u =
O_{k-4}(r^{-1})\;\;,\;r=\sqrt{\rho^2+z^2}\rightarrow\infty\;. \ee

\section{Reduction to a harmonic map problem}
\label{Sharmonic}

  The electro-vacuum field equations~\eqref{fieldeq1}-\eqref{fieldeq3} and simple-connectedness of $\doc$  guarantee the global existence of the
following potentials:
\newcommand{\newomega}{v}
\bel{fields} d\chi=i_{K_\kl 1}F\;\;,\;\;\;
 d\psi=i_{K_\kl 1}*F\;\;\text{ and }\;\; d\newomega=\omega-2(\chi d\psi-\psi
d\chi)\;,
\ee
where
\bel{twist potential} \omega:=*(dK_\kl 1^{\flat}\wedge K_\kl
1^{\flat})\;, \ee
is the axial twist form. As discussed in detail
in~\cite{Weinstein96}, when a global representation
in terms of Weyl coordinates like~\eqref{space-time metric} is
allowed, the space-time metric is uniquely determined by an
axisymmetric harmonic map
\bel{harmonic map}
 \Phi=(\lambda,\newomega,\chi,\psi):\R^3\setminus \mcA
 \longrightarrow \mathbb{H}^2_{\mathbb{C}}\;,
\ee
here $\mcA=\{(0,0,z)\ | \ z\in\R\}$ and $\mathbb{H}^2_{\mathbb{C}}$ is the `upper half-space
model' of the 2-dimensional complex hyperbolic space, i.e., $\R^4$
with metric given by
\bel{upper-half metric} ds^2=d\lambda^2+e^{4\lambda}(d\newomega+\chi
d\psi-\psi d\chi)^2+e^{2\lambda}(d\chi^2+d\psi^2)\;. \ee
%.

The metric coefficient $\lambda$ is part of the harmonic map and the remaining unknowns of the metric
can be determined from $\Phi$
by considering the unique solution $(w,u)$ of the set of
equations
\bel{Ernst} \partial_\rho w  = -e^{4\llambda}\rho\:\omega_z
 \;, \qquad
 \partial_z w = e^{4\llambda}\rho\:\omega_{\rho}
 \;,
\ee
\bel{Ernst2}
\partial_{\rho} u -\partial_{\rho}\llambda
 = \rho\left[ (\partial_{\rho}\llambda)^2 -(\partial_{z}\llambda)^2
+\frac{1}{4}e^{4\llambda}(\omega_{\rho}^2-\omega_z^2)
+e^{2\lambda}\left((\partial_{\rho}\chi)^2 -(\partial_{z}\chi)^2
+(\partial_{\rho}\psi)^2-(\partial_{z}\psi)^2 \right) \right] \ee
\bel{Ernst3}
\partial_{z} u - \partial_{z}\llambda= 2\:\rho\left[ \partial_{\rho}\llambda\: \partial_{z}\llambda
+\frac{1}{4}e^{4\llambda}\omega_{\rho}\:\omega_z
+e^{2\lambda}(\partial_{\rho}\chi\: \partial_{z}\chi
+\partial_{\rho}\psi\:\partial_{z}\psi) \right] \;, \ee
that go to zero at infinity, and where we write $\omega_a:=\omega(\partial_a)$ for
$a\in\{\rho, z\}$.

\subsection{Distance function on the target manifold}
\label{Sdistance}

The criteria for uniqueness of harmonic maps
used in this paper (see Theorem~\ref{Tuniquehm} and compare~\cite[Appendix
C]{CLW}), is stated in terms of the pointwise distance  between the maps.
For the `disk model' of $\mathbb{H}^2_{\mathbb{C}}$ the
distance between two points $z=(z_1,z_2)$ and $w=(w_1,w_2)$ is given
by~\cite[eq 55, pg 26]{Weinstein94}
\bel{dist disk}
  \cosh(d) =\,\frac{\abs{ 1-\bar z_1\,w_1-\bar z_2\,w_2 } }{ \sqrt{1-\abs{z}^2}\sqrt{1-\abs{w}^2}
  }\;.
\ee
To obtain the distance function for the `upper half-space
model' we will use the isometry
between the two referred  models presented in~\cite[Appendix]{Weinstein94}: first we perform the coordinate transformation
 $$z_1=\frac{1-x_1}{1+x_1}\;\;\;,\;\;\; z_2=\frac{2x_2}{1+x_1}\;,$$
with analogous expressions for $w_i=w_i(y_1,y_2)$ to obtain
$$\abs{ 1-\bar z_1\,w_1-\bar z_2\,w_2 }=\frac{2 \abs{\bar x_1
+ y_1-2\,\bar x_2\, y_2}}{\abs{ 1+x_1 } \abs{ 1+y_1 } }\;; $$
then we take
$$e^{\lambda_1}=\frac{ \abs{ 1+z_1 } }{\sqrt{1-\abs{z}^2}}\;\;\;\text{ and }\;\;e^{\lambda_2}=\frac{\abs{1+h_1}}{\sqrt{1-\abs{h}^2}}$$
so that
\bel{distance 3} \cosh(d) =\frac{1}{2} \abs{ \bar x_1 + y_1-2\,\bar x_2\,
y_2} e^{\lambda_1+\lambda_2}\;; \ee
and finally, by writing
\bel{} x_1=e^{-2\lambda_1}+\chi_1^2+\psi_1^2+2\,i\newomega_1
\;\;\;\text{ and }\;\; x_2=\chi_1+i\psi_1\;, \ee
with similar expressions for
$y_i=y_i(\lambda_2,\newomega_2,\chi_2,\psi_2)$, we see that  the
distance function satisfies
\footnote{ By taking $\chi_i=\psi_i\equiv 0$ we see that this
distance function is related to the one used in the vacuum
case~\cite[Section 6.5.1]{ChrusCosta} by $d=2 d_b$. This discrepancy has its
genesis in an analogous relation between the line elements of the
different disk models used.}
\begin{align}
\label{dist EM}
  \cosh^2(d)& =\,\frac{1}{4}\, e^{2(\lambda_1+\lambda_2)}(e^{-2\lambda_1}+ e^{-2\lambda_2}+ (\chi_1-\chi_2)^2+ (\psi_1-\psi_2)^2)^2 \\
   \nonumber         & +  e^{2(\lambda_1+\lambda_2)}
(\newomega_2-\newomega_1 - \chi_1\psi_2 + \chi_2\psi_1)^2
\\
\nonumber
&=\,\frac{1}{4}\left\{
e^{-\lambda_1+\lambda_2}+ e^{\lambda_1-\lambda_2}+ e^{\lambda_1+\lambda_2} (\chi_1-\chi_2)^2+e^{\lambda_1+\lambda_2} (\psi_1-\psi_2)^2
\right\}^2
 \\
   \nonumber         & +
e^{2(\lambda_1+\lambda_2)}\left\{(\newomega_2-\newomega_1)+(\chi_2\psi_1
- \chi_1\psi_2) \right\}^2 \;,
\end{align}
or in an apparently more intrinsic way
\begin{align}
\cosh^2(d)& =\,\frac{1}{4}\left\{
 \sqrt{ \frac{ \fourg_2(\partial_{\varphi},\partial_{\varphi})}{\fourg_1(\partial_{\varphi},\partial_{\varphi})} }+
\sqrt{ \frac{ \fourg_1(\partial_{\varphi},\partial_{\varphi})}{\fourg_2(\partial_{\varphi},\partial_{\varphi})} }+
\frac{(\chi_1-\chi_2)^2+(\psi_1-\psi_2)^2}{\sqrt{\fourg_1(\partial_{\varphi},\partial_{\varphi})} \sqrt{\fourg_2(\partial_{\varphi},\partial_{\varphi})} }
\right\}^2
 \\
   \nonumber         & +
\left\{ \frac{(\newomega_2-\newomega_1)+(\chi_2\psi_1 - \chi_1\psi_2
) } {\sqrt{\fourg_1( \partial_{\varphi},\partial_{\varphi}) } \sqrt{
\fourg_2(\partial_{\varphi},\partial_{\varphi})} } \right\}^2 \;.
\end{align}

 It will also be helpful to use the usual rescaling
\bel{U}
  U_i=\lambda_i+\ln\rho\;, \quad \mbox{so that} \quad
 \fourg_i(\partial_\varphi,\partial_\varphi) = \rho^2 e^{-2U_i}= e^{-2\lambda_i}
 \;,
  \ee
from which we get our final expression for the distance in the
`upper half-space':
\begin{align}
\label{dist}
  \cosh^2(d) =\,&\frac{1}{4}\left\{e^{U_1-U_2}+e^{-U_1+U_2} +\rho^{-2}e^{U_1+U_2} (\chi_1-\chi_2)^2+ \rho^{-2}e^{U_1+U_2} (\psi_1-\psi_2)^2\right\}^2 \\
   \nonumber         & + \left\{ \rho^{-2}e^{U_1+U_2}(
\newomega_2-\newomega_1) - \rho^{-2}e^{U_1+U_2}(\chi_1\psi_2 -
\chi_2\psi_1)\right\}^2\;.
\end{align}

\section{Boundary conditions}
\label{sBoundary}

\subsection{The Axis}
\label{Saxis}

 From now on we will be controlling the distance, as given by any of the formulae in the previous section,
 between the harmonic maps arising
from two \regular, stationary-axisymmetric and electro-vacuum
space-times $(\mcM_i,\fourg_i)$, $i=1,2$. We will start by showing
that
\bel{goal axis}
d(\Phi_1,\Phi_2)\;\;\text{is bounded near } \overline{\mcA\cap\doc}
\;.
\ee
 In this section we will be working with the following
coordinate systems: isothermal coordinates $(\hat x_i,\hat z_i)$
globally defined in the doubling across the axis  of the orbit space
of an appropriate extension of the $\U(1)$ action to the manifold
obtained by the addition of 3-discs to every connected component of
$\partial\hyp_i$;\footnote{The resulting space is diffeomorphic to
$\R^2$, see Figure~\ref{Fquot}, and for more details concerning this
construction see~\cite[Section 6]{ChrusCosta}; also, the fact that
$I^+$-regularity, stationarity and the null energy condition imply
spherical topology for the connected components of the cross-section
of the event horizon $\partial\hyp_i$ follows from~\cite{ChWald}.}
``canonical coordinates" $(\rho, z)$ of the half plane $\R^+_0\times
\R$, which is the image of each (physical) orbit space by the  map
$\Psi_i$ defined by $(\hat x_i,\hat z_i)\mapsto(\rho_i(\hat x_i,\hat
z_i),z_i(\hat x_i,\hat z_i))$.

\begin{figure}[t]
\begin{center} {
\resizebox{3.5in}{!}{\includegraphics{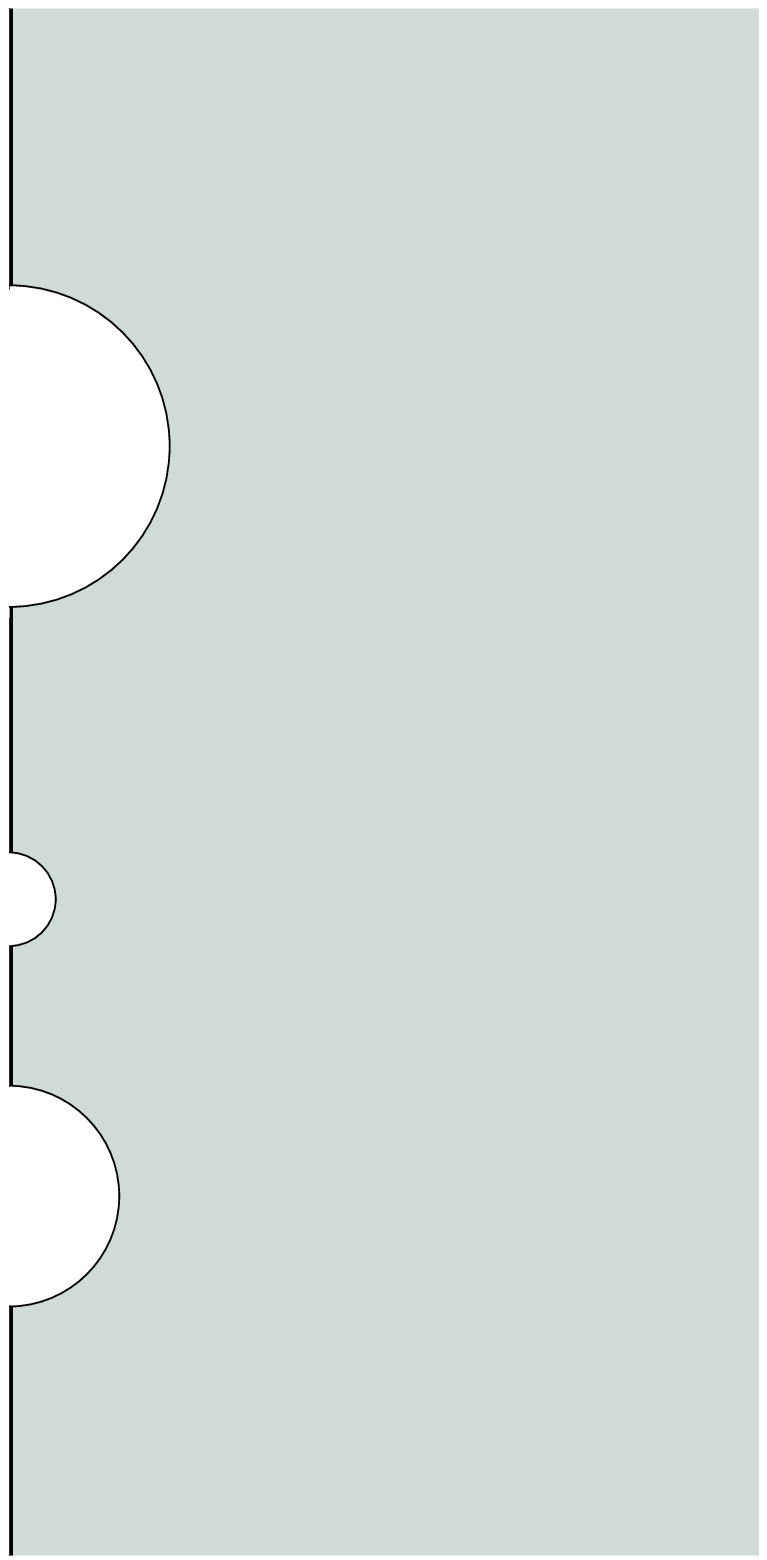}
\includegraphics{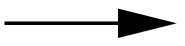}\includegraphics{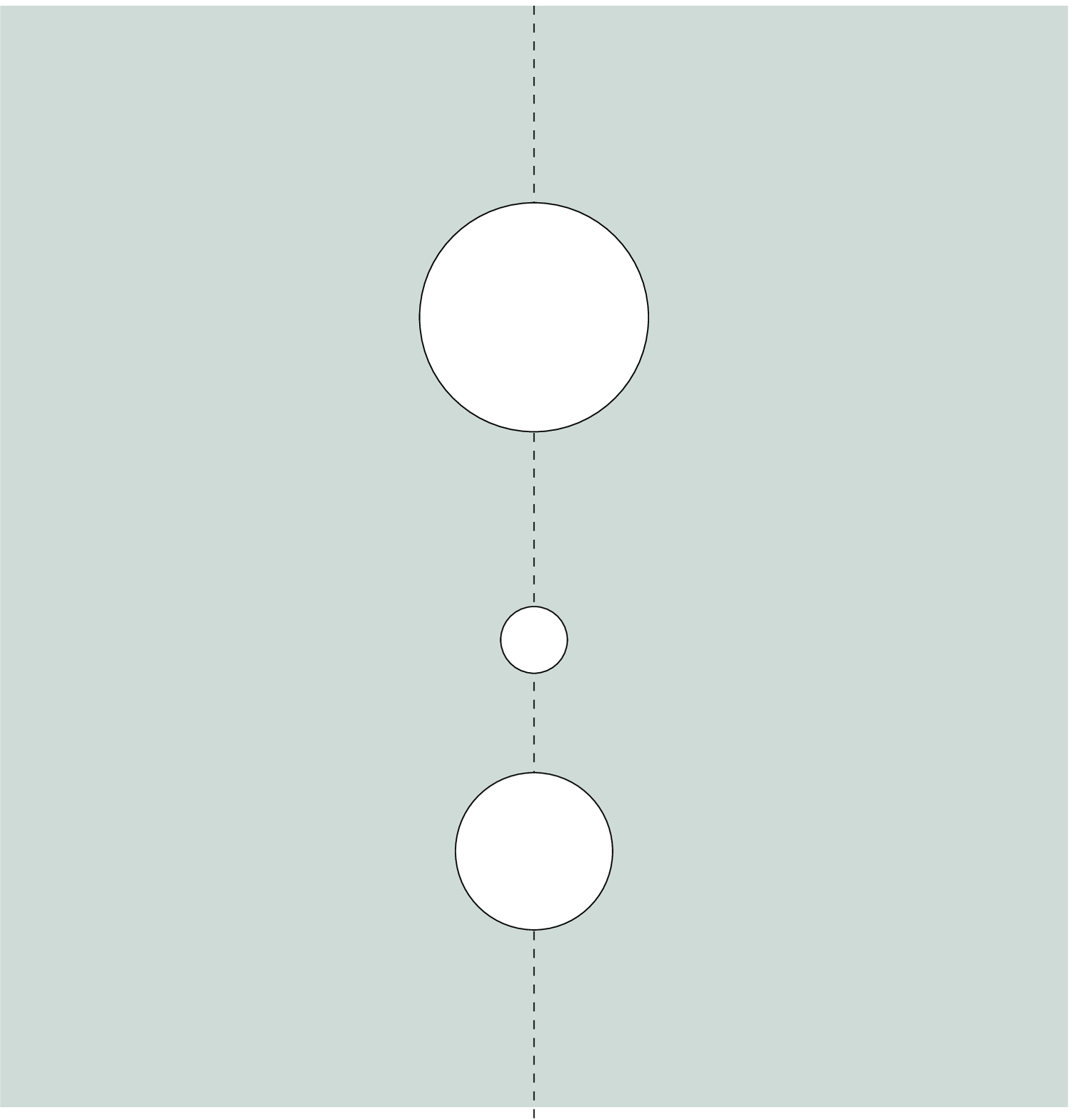}}
}
\caption{The quotient space and its double.
\label{Fquot}}
\end{center}
\end{figure}

Let $\phi_s$ be the flow generated by the axial Killing vector
$K_\kl 1$. In the doubling of the orbit space the isothermal
coordinates satisfy
$$  \hat x \circ \phi_{\pi} = -\hat x \;\;\;\; \text{ and }\;\;\; \hat z  \circ \phi_{\pi}  = \hat z\;. $$
Then, invariance of a function $(\hat x, \hat z)\mapsto f(\hat
x,\hat z)$ under the axial flow, which is the case for the fields
$\newomega, \chi$ and $\psi$, implies that the function $\hat
x\mapsto f(\hat x, \hat z)$ is even for all $\hat z$. In this case,
if $f$ is $C^2$,  Taylor expanding on $\hat x$, from the axis, gives
\bel{even expansion} f(\hat x, \hat z)=f(0, \hat
z)+\frac{1}{2}\frac{\partial^2f}{\partial \hat x^2}(c(\hat x),\hat
z)\,\hat x^{\,2}\;\;,\;  \abs{c(\hat x)}\leq \abs{\hat x}\;. \ee
Now fix a point in $\overline{\mcA\cap\doc}$ and by rescaling $\hat
z$ assume it lies at the origin. Suppose also that  $f\equiv
f_0:=f(0,0)$  along the connected component of $\overline{\mcA\cap
\doc}$, in $\overline\doc$, containing $(0,0)$; this is clearly the
case  for  all the functions  appearing in~\ref{dist EM} and it also
implies that we can realize the aforementioned extension of
the doubling of the orbit space to $\R^2$ while preserving the
constancy of $f$ along the extended axis near the poles, i.e., near
the points where the axis meets the event horizon. Then~\eqref{even
expansion} implies
\bel{even control} \abs{ f(\hat x, \hat z)-f_0 }\leq C \hat
x^{\,2}\;\;\text{ near } (0, 0)\in \overline{\mcA\cap\doc}\;. \ee
We will need bigger control over the functions
$e^{-2\lambda}=\fourg_{\varphi\varphi}$. To this end let $\{x,y,z\}$
be Gaussian coordinates along the axis, in the extension of $\hyp$,
with $\mcA=\{x=y=0\}$ and for which $K_\kl
1=x\partial_y-y\partial_x$ (see~\cite[pg 5]{ChUone} and compare
with~\eqref{eta}). For any path with initial velocity transverse to
$\mcA$ we have
\bel{trans} \nabla_{\dot \gamma(0)}K_\kl
1|_{x=y=0}=\nabla_{\gamma^i\partial_i}(x\partial_y-y\partial_x)|_{x=y=0}=\gamma^x\partial_y-\gamma^y\partial_x\;,
\ee
and consequently $\fourg(\nabla_{\dot \gamma(0)}K_\kl 1,\nabla_{\dot
\gamma(0)}K_\kl 1)=(\gamma^x)^2+(\gamma^y)^2\neq 0$.   Since
$\nabla_{\mu}\fourg(K_\kl 1,K_\kl 1)=2\,\fourg(\nabla_{\mu}K_\kl
1,K_\kl 1)$  we see that the gradient of $\fourg_{\varphi\varphi}$
vanishes at the axis and
$$\nabla_{\mu}\nabla_{\nu\,}\fourg(K_\kl 1,K_\kl 1)|_{\mcA}=2\,\fourg(\nabla_{\mu} K_\kl 1, \nabla_{\nu} K_\kl 1)|_{\mcA}\;.$$
  Taylor expanding along $\gamma$ yields
$$ \fourg_{\varphi\varphi}\circ \gamma(s) = (\underbrace{\fourg (\nabla_{\dot \gamma(0)} K_\kl 1, \nabla_{\dot \gamma(0)} K_\kl 1)}_{\neq 0}+O(s))s^2\;,$$
from which it follows that {\em for any path transverse to $\mcA$ and small $s$}
\bel{control u}
C^{-1}\, s^2\leq \fourg_{\varphi\varphi} \circ \gamma(s)\leq C\,s^2\;.
\ee

\bigskip

We will need to consider two separate cases. First, fix, in each
space-time, a point belonging to $\mcA_i\cap\doc$ and rescale all
the previous coordinate systems so that each of the fixed points
corresponds to its respective origin and $\Psi_i(0,0)=(0,0)$. At
these points, since there the boundary of the orbit space is analytic, the
function $\rho_i=\rho_i(\hat x_i,\hat z_i)$ may be extended
analytically  across the origin, therefore, as an immediate
consequence of~\eqref{control u} we get control over the first terms
appearing in~\eqref{dist}
\bel{control U}
 e^{U_j-U_i}=\sqrt{\frac{\fourg_i(\partial_\varphi,\partial_\varphi) }{\fourg_j(\partial_\varphi,\partial_\varphi)}}
 \le \sqrt{\frac{C_i \rho^2 }{C_j^{-1}\rho^2}}\leq C\;\;\text{ near } (0,
0)\in\mcA\cap\doc\;.
 \ee
Since the $\chi_i$'s and the $\psi_i$'s are all bounded near the
origin our goal gets reduced to showing that
\bel{goal axis2} \rho^{-2}e^{U_1+U_2}(f_1-f_2)=O(1)\text{ near } (0,
0)\in\mcA\cap\doc\;, \ee
when $f_1=\chi_1,\psi_1,\newomega_1,\chi_1\psi_2$  and
$f_2=\chi_2,\psi_2,\newomega_2,\chi_2\psi_1$, where by this we mean
that if, for example, we set $f_1=\chi_1$ then $f_2=\chi_2$.

Let us start with $f_1=\chi_1,\psi_1,\newomega_1$  and
$f_2=\chi_2,\psi_2,\newomega_2$. Each $f_i$ is invariant under the
respective axial flow and constant along each connected component of
$\mcA_i\cap \doc$, so if we impose $f_1(0,0)=f_2(0,0)=f_0$, which is
always achievable if the space-times $(\mcM_i,\fourg_i)$ have the
same set of masses, angular momenta and charges (see
Section~\ref{sWeinstein} and~\cite[Section 2.3]{Weinstein96}), we
see that~\eqref{even control} holds and using~\eqref{control u}
and~\eqref{control U}  we get
\beal{omecont}
&&\\
\nonumber
\abs{\rho^{-2}e^{U_1+U_2}(f_1-f_2)}
 & = &
\abs{\frac {f_1-f_2}{ \sqrt{\fourg_1(\partial_\varphi,\partial_\varphi)}\sqrt{ \fourg_2(\partial_\varphi,\partial_\varphi) }}}
 \le
\frac {\abs{f_1- f_0}+\abs{f_2-f_0}}{ \sqrt{\fourg_1(\partial_\varphi,\partial_\varphi)}\sqrt{ \fourg_2(\partial_\varphi,\partial_\varphi) }}
 \\
\nonumber
 & = &
\frac {\abs{ f_1- f_0 }}{ \fourg_1(\partial_\varphi,\partial_\varphi) }
 {\sqrt{\frac { \fourg_1(\partial_\varphi,\partial_\varphi)} { \fourg_2(\partial_\varphi,\partial_\varphi)} } }+
 \frac {\abs{ f_2- f_0 }}{ \fourg_2(\partial_\varphi,\partial_\varphi) }
 { \sqrt{ \frac { \fourg_2(\partial_\varphi,\partial_\varphi)} { \fourg_1(\partial_\varphi,\partial_\varphi)} } }
 \\
 \nonumber
 & \le & \frac{C_1 \hat x_1^2}{C_2^{-1} \hat x_1^2 } C_3 +\frac{C_4 \hat x_2^2}{C_5^{-1} \hat x_2^2 } C_6
\\
\nonumber
& \le & C
 \;\;\text{ near } (0, 0)\in\mcA\cap\doc
 \;.
\eean
We take the chance to stress the fact that the previous argument
does not apply to the fields $\chi_1\psi_2$ and $\chi_2\psi_1$ since
these products involve functions originating from different
space-times and therefore only make sense as functions of $(\rho,z)$
for which estimates like~\ref{even control} are not available a
priori.\footnote{In fact, extending $\rho_i$ and $z_i$ near this axis
points by $\rho_i(-\hat x_i,\hat z_i)=-\rho_i(\hat x_i,\hat z_i)$
and $z_i(-\hat x_i,\hat z_i)=z_i(\hat x_i,\hat z_i)$ shows that
invariance under the axial flow implies that
$f(\rho,z):=f\circ\Psi_i^{-1}(\rho,z)$ is an even function of
$\rho$. Then, direct estimates in terms of $\rho$ analogous
to~\eqref{even control} may be obtained for all the fields and the presented
procedure including~\eqref{omecont} may be bypassed. Unfortunately
this is no longer possible near points where the axis meets the
horizon as the $\rho_i$ are no longer differentiable.}

%an even function of $\hat x_i$ then
%$f(\rho,z):=f\circ\Psi_i^{-1}(\rho,z)$ is an even function of
%$\rho$. As an immediate consequences we get (compare~\eqref{even
%control}) %%
%\bel{even control 2} \abs{ f(\rho, z)-f_0 }\leq C
%\rho^{\,2}\;\;\text{ near } (0, 0)\in \overline{\mcA\cap\doc}\;. \ee
%%

To bypass this problem we write
$$ \chi_1\psi_2 -
\chi_2\psi_1 =(\chi_1+\chi_2)(\psi_2-\psi_1)+\chi_1\psi_1-\chi_2\psi_2\;.$$
Since $\chi_1+\chi_2$ is bounded, to control the first term we just need to take $f_i=\psi_i$ as before.
Setting $f_1=\chi_1\psi_1$ and $f_2=\chi_2\psi_2$ we see that the previous
argument still applies as these are also axially symmetric functions which
are constant along the axis components. The desired result follows.

\bigskip

To finish the proof of boundedness of~\eqref{dist} near the singular
set $\mcA\cap\doc$ we still have to analyze what happens near points
where the axis meets the horizon. Choose such a point in each
space-time and, without loss of generality, assume that these are `north
poles' which, as before, lie at the origin of the coordinate systems
$(\hat x_i, \hat z_i)$, and satisfy  $(\rho, z)=\Psi_i(0,0)=(0,0)$.

As already mentioned, a careful extension of the doubling of the
orbits spaces validates~\eqref{even control} in a neighborhood of
these `north poles', but, on the other hand, the $\rho_i$'s are now
non-differentiable at such points and~\eqref{control U} no longer
holds. Nonetheless, if we are able to control $e^{U_i-U_j}$ by
other means, then the inequalities established in~\eqref{omecont}
extend to the case under consideration and boundedness of the
distance near the axis follows. This problem, which is in fact the
major difficulty that arises in the analysis of the boundary
conditions of this axisymmetric harmonic maps, has been recently
overcomed for the vacuum case~\cite[Section 6.5.1]{ChrusCosta} by
obtaining the following uniform estimate
\bel{UbbhvOLD}
 U =   \ln \sqrt{z+\sqrt{z^2+\rho^2}} +  O(1) \ \mbox{ near
}(0,0)\in\mcA\cap\mcEp
 \;,
\ee
from which the desired consequence immediately follows. This result,
which requires this component of the horizon to be non-degenerate,
extends to the electro-vacuum case immediately.

\subsection{Spatial infinity}
\label{Sinfinity}

 In this section we want to show that
\bel{goal infty}
\lim_{\sqrt{\rho^2+z^2}\rightarrow+\infty} d(\Phi_1,\Phi_2)=0
\;,
\ee
with $d$ implicitly defined by~\eqref{dist}. For this we will assume
stationarity and asymptotic flatness as given by the system of
equations~\eqref{gme2}--\eqref{foff A}. It turns out that the
estimates provided by asymptotic flatness, even in the way just
defined, seem insufficient to control the relevant fields; even in
an adapted frame provided by the results of Section~\ref{sWeyl}
integration of the defining equations~\eqref{fields} yields
divergent logarithmic terms. Fortunately, in the stationary and
electro-vacuum setting, the asymptotic analysis of Beig and
Simon~\cite{BeigSimon,Simon} provides  relevant improvements of the
initial decay rates by means of the expansion~\eqref{asympt}.

Let $D$ and $\epsilon_{ijk}$
denote the covariant derivative and volume element of $\gamma$, the
induced metric in $\hyp_{ext}$ as in Section~\ref{Sprelim}. A well
known consequence of the source free Einstein-Maxwell
equations~\eqref{fieldeq1}-\eqref{fieldeq3} and simple-connectedness
of $\doc$  is the global existence of functions $\tau$ and $\sigma$
satisfying~\cite{IsraelWilson, Simon, Weinstein96}
\bel{tau_i} D_i\tau=V^2\epsilon_i{}^{jk}(D_jA_k+\theta_j D_kA_0)\;,
\ee
and
\bel{sigma_i} D_i\sigma =-V^4\epsilon_i{}^{jk}D_j\theta_k +i( \bar{
\Psi} \partial_i \Psi - \Psi \partial_i\bar \Psi)\;, \ee
where $\Psi:=A_0+i\tau$. If we  introduce the Ernst potential
\bel{Ernst pot}
\mycal{C}=V^2-\Psi\bar\Psi+i\sigma\;,
\ee
and consider the complex valued fields
$\newv$ and $\neww$,\jlca{$v$
and $w$ changed to $\newv$ and $\neww$ from here on! Needs
checking.} implicitly defined by
\bel{w and v} \mycal{C}=\frac{1-\neww}{1+\neww}\;\;\;\;\; ,
\;\;\;\;\;\Psi=\frac{\newv}{1+\neww}\;, \ee
then~\cite[eq 3.11]{Simon} provides the following expansion for the
vector $\mycal{E}^A:=(\neww,\newv)\in \C^2$ in terms of an arbitrary
asymptoticaly flat coordinate system
\bel{asympt}
\mycal{E}^A=\frac{M^A}{r}+\frac{M_k^Ax^k}{r^3}+O_{\infty}(\log
r/r^3)\;. \ee
We note that the apparent discrepancy between the error term here
with the one in the original paper comes from the fact that the
result there is presented in adapted coordinates  obtained from
arbitrary asymptotically flat coordinates by a transformation of the
form $x^i\mapsto x^i+O_{\infty}(\log r)$.

 Using the identity $\frac{A}{B+C}=\frac{A}{B}-\frac{AC}{B(B+C)}$ we get
\begin{align*}
\mycal{C}& =
\frac{1-\neww}{1+\neww}=\frac{1+\neww-2\:\neww}{1+\neww} =
1-\frac{2\:\neww}{1+\neww}=1-2\left(\neww-\frac{\neww^2}{1+\neww}
\right)
\\
&= 1-2\left(\neww-\neww^2+\frac{\neww^3}{1+\neww}  \right) \;.%}
\end{align*}
 Inserting the $\neww$-component of~\eqref{asympt} into the last expression yields
\begin{align*}
\mycal{C}& = 1-2\left( \frac{M^{\neww}}{r} +
\frac{M_k^{\neww}x^k}{r^3}+O_{\infty}(\log r/r^3) \right) +2\left(
\frac{M^{\neww}}{r}+\frac{M_k^{\neww}x^k}{r^3}+O_{\infty}(\log
r/r^3) \right)^2 +O_{\infty}(r^{-3})
\\
&=
1-2\frac{M^{\neww}}{r}-2\frac{M_k^{\neww}x^k}{r^3}+2\frac{(M^{\neww})^2}{r^2}+O_{\infty}(r^{-4})+O_{\infty}(\log
r/r^3) \;.
\end{align*}
Noting that  the topological restrictions imposed by asymptotic
flatness imply that the imaginary part of $M^{\neww}$ vanishes, $\Im
M^{\neww}=0$~\cite[Section IV]{Simon}, we write $M^{\neww}=M$ and by
setting $M_k^{\neww}=M_k+iS_k$ we get
\bel{mcC}
\mcC=1-2\frac{M}{r}+2\frac{M^2}{r^2}-2\frac{M_kx^k}{r^3}-2\:i\frac{S_kx^k}{r^3}+O_{\infty}(\log
r/r^3) \;. \ee
Consequently
\bel{sigma} \sigma=\Im
\mycal{C}=-2\frac{S_kx^k}{r^3}+O_{\infty}(\log r/r^3) \;. \ee

 Similarly for $\Psi$ we get
\begin{equation*}
\Psi  = \frac{\newv}{1+\neww}=\newv-\frac{\newv \neww}{1+\neww} =
\newv-\newv \neww+\frac{\newv \neww^2}{1+\neww}\;.
\end{equation*}
Inserting~\eqref{asympt} into the last expression yields
\begin{align*}
\Psi & = \newv(1-\neww)+O_{\infty}(r^{-3})
\\
&= \left(
\frac{M^{\newv}}{r}+\frac{M_k^{\newv}x^k}{r^3}+O_{\infty}(\log
r/r^3) \right) \left(
1-\frac{M^{\neww}}{r}-\frac{M_k^{\neww}x^k}{r^3}+O_{\infty}(\log
r/r^3) \right) +O_{\infty}(r^{-3})
\\
&=
\frac{M^{\newv}}{r}-\frac{M^{\newv}M^{\neww}}{r^2}+\frac{M_k^{\newv}x^k}{r^3}+O_{\infty}(\log
r/r^3) \;.
\end{align*}
As before $\Im M^v=0$. So now, by setting  $M^{\newv}=\frac{Q}{2}$
and $M_k^\newv=Q_k+iB_k$, we see that
\bel{A_0} A_0=\Re \Psi =
\frac{Q}{2\:r}-\frac{MQ}{2\:r^2}+\frac{Q_kx^k}{r^3}+O_{\infty}(\log
r/r^3)\;, \ee
\bel{tau} \tau=\Im \Psi = \frac{B_kx^k}{r^3}+O_{\infty}(\log
r/r^3)\;. \ee

We have $\bar \Psi \partial_i \Psi - \Psi \partial_i\bar \Psi=2i(A_0
\partial_i\tau-\tau \partial_iA_0)=O_{\infty}(r^{-4})$
 and using~\eqref{foff} and~\eqref{sigma} we get
\bel{} \epsilon_i{}^{jk} D_j\theta_k = -V^{-4}\left(D_i\sigma -i(
\bar \Psi D_i \Psi - \Psi D_i\bar \Psi)\right)=
D_i\left(2\frac{S_kx^k}{r^3}+O_{\infty}(\log r/r^{3})\right)\;. \ee
 With the exception of the already noted $\log r$  discrepancy in the error term, this is~\cite[eq 4.1, pg 1010]{BeigSimon} and so we get
\jlca{consider (understanding better and) doing the full computation}
\bel{teta}
\theta_i:=\frac{\fourg_{it}}{\fourg_{tt}}=2e_{ijk}\frac{S^j x^k}{r^3}+O_k(\log r/r^3)\;,
\ee
where $e_{[ijk]}=e_{ijk}$ with $e_{123}=1$.

\subsubsection{The electromagnetic twist potential and the norm of the axial Killing vector}
\label{Stwist}

Until now we have been working with a generic asymptotically flat
coordinate system, but to estimate the electromagnetic twist
potential $\newomega$ via the Ernst equations~\eqref{Ernst} and the
results of the previous section we will need to use adapted
coordinates. %that will allow us to control the remaining terms in
%this equations, including the norm of the axial Killing vector,
%beyond the, already alluded, insufficient information provided a
%priori by asymptotic flatness.
 So, let $\{t,\varphi, \rho, z\}$ be
the Weyl coordinates as constructed in Section~\ref{sWeyl} and
define the cylindrical type coordinates
 \bel{xyz}
 \left\{
 \begin{array}{l}
 x=\rho\cos\varphi \\
 y=\rho\sin\varphi
\end{array}\right.\;.
\ee

 A simple but noteworthy fact is that in this coordinate system we have
 \bel{eta}
 K_\kl 1=\partial_{\varphi}=x\partial_y-y\partial_x\;.
\ee
The estimates of the previous section will only be available to us
in this coordinates if $\{t,x^i\}=\{t,x, y, z\}$ is an
asymptotically flat coordinate system. This is in fact the case. To
see it note that in the orbit space $\{t=\varphi=0\}$ the
identity~\cite[eq 6.9]{ChrusCosta}
 yields
\bel{partial rho}
\partial_{\rho}=(1+O_k(\hat r^{-1}))\partial_{\hat x}+O_{\infty}(\hat r^{-1})\partial_{\hat z}\;\;\;\; (\varphi=0)\;,
  \ee
\bel{partial z}
\partial_{z}=O_{\infty}(\hat r^{-1})\partial_{\hat x}+(1+O_{\infty}(\hat r^{-1}))\partial_{\hat z}\;\;\;\;\;
(\varphi=0)\;.
 \ee
Recall that $\{\hat x, \hat z\}$ are asymptotically flat isothermal
coordinates (for the orbit space metric). Direct computations yield
$\fourg_{xx}|_{\varphi=0}=1+O_{\infty}(r^{-1})$, a similar
expression for $\fourg_{yy}$ and, using
$\fourg_{\rho\varphi}\equiv0$, also  $\fourg_{xy}|_{\varphi=0}=0$.
The defining decay rates are now obtained by flowing the previous
estimates along the integral lines of the axial Killing vector. We
illustrate this with an explicit calculation:
\begin{align*}
  \fourg_{xy}|_{\varphi=-\varphi_0}&=\fourg((\phi_{\varphi_0})_*\partial_x,(\phi_{\varphi_0})_*\partial_y)|_{\varphi=0}             \\
          &=\fourg(\cos\varphi_0\partial x +\sin\varphi_0\partial y,-\sin\varphi_0\partial x +\cos\varphi_0\partial y)|_{\varphi=0} \\
          &=-\sin\varphi_0\cos\varphi_0 \underbrace{\fourg_{xx}|_{\varphi=0}}_{=1+O_{\infty}(r^{-1})}
          (\cos^2\varphi_0-\sin^2\varphi_0)\underbrace{\fourg_{xy}|_{\varphi=0}}_{=0}+\sin\varphi_0\cos\varphi_0
          \underbrace{\fourg_{yy}|_{\varphi=0}}_{=1+O_{\infty}(r^{-1})}\\
          &=O_{\infty}(r^{-1})
          \;.
\end{align*}

 So we have constructed asymptotically flat coordinates for which
 the following uniform estimate holds
 \bel{estimate u}
\fourg_{\varphi\varphi}|_{\varphi=0}=\rho^2\fourg_{yy}|_{\varphi=0}=\rho^2(1+O_{\infty}(r^{-1}))\;.
\ee
As a nice consequence we get
 \bel{estimate U}
e^{-2U}:=\frac{\fourg_{\varphi\varphi}}{\rho^2}=1+O_{\infty}(r^{-1})\;,
\ee
from which we see that
 \bel{estimate Ui - Uj}
e^{U_i\pm U_j}:=(1+O_{\infty}(r^{-1}))(1+O_{\infty}(r^{-1}))^{\pm
1}=1+O_{\infty}(r^{-1})\rightarrow_{r\rightarrow+\infty} 1\;, \ee
and our goal~\eqref{goal infty} gets reduced to showing that
\begin{align}
\label{goal2}
\lim_{\sqrt{\rho^2+z^2}\rightarrow +\infty} \frac{(\psi_1-\psi_2)^2}{\rho^{2}}
&=
\lim_{\sqrt{\rho^2+z^2}\rightarrow +\infty} \frac{(\chi_1-\chi_2)^2}{\rho^{2}}
\\
\nonumber &= \lim_{\sqrt{\rho^2+z^2}\rightarrow +\infty}
\frac{\newomega_1-\newomega_2}{\rho^{2}}
\\
\nonumber
&=
\lim_{\sqrt{\rho^2+z^2}\rightarrow +\infty} \frac{\chi_1\psi_2-\chi_2\psi_1}{\rho^{2}}
=0
\;.
\end{align}

\bigskip

It follows from~\eqref{teta} and~\eqref{space-time metric} that
 $$\fourg_{zt}\equiv 0\Rightarrow\theta_z\equiv0\Rightarrow S_x=S_y=0\;.$$
So we set $J:=-S_z$ and by using~\eqref{teta} with~\eqref{foff} we
get
\bel{teta2}
\fourg_{yt}|_{\varphi=0}=2J\frac{\rho}{r^3}+O_{\infty}(\log
r/r^3)\;, \ee
from which
\bel{g phi t} \fourg_{\varphi
t}|_{\varphi=0}=\fourg(\rho\partial_y,\partial_t)|_{\varphi=0}=2J\,\frac{\rho^2}{r^3}+
\rho\, O_{\infty}(\log r/r^3)\;, \ee
and therefore
\bel{}
 \frac{\fourg_{\varphi t}}{\fourg_{\varphi \varphi}}|_{\varphi=0}= \frac{2J}{r^3}+\frac{1}{\rho}O_{\infty}(\log r/r^3)
          \;.
 \ee
%

%%
%\begin{align*}
%  \frac{\fourg_{\varphi t}}{\fourg_{\varphi \varphi}}|_{\varphi=0}&= \frac{-2S\frac{\rho^2}{r^3}+ \rho O_{\infty}(\log r/r^3)}{\rho^2(1+O_{\infty}(r^{-1}))}
%           =\frac{-2S}{r^3+O_{\infty}(r^{2})}+\frac{1}{\rho}O_{\infty}(\log r/r^3) \\
%           &= -2S\left(\frac{1}{r^3}+O_{\infty}(r^{2-6})\right)+\frac{1}{\rho}O_{\infty}(\log r/r^3)\\
%          &= \frac{-2S}{r^3}+\frac{1}{\rho}O_{\infty}(\log r/r^3)
%          \;\,
%\end{align*}
%%
%where we have used the identity
%%
%\jlca{this as been used implicitly in previous occasions}
%%
%\begin{Lemma}
%%
%$$ \frac{A}{r^k+O_{\infty}(r^l)}=\frac{A}{r^k}+A\,O_{\infty}(r^{\,l-2k})\;\;\text{ for }k>l\;.$$
%%
%\end{Lemma}
%
%\begin{proof}
%The  general identity $\frac{A}{B+C}=\frac{A}{B}-\frac{AC}{B(B+C)}$, leads to
%%
%$$ \frac{A}{r^k+O_{\infty}(r^l)}=\frac{A}{r^k}+\frac{AO_{\infty}(r^{l-k})}{r^k+O_{\infty}(r^l)}\;.$$
%%
%For $k>l$ and $r$ large enough $|O_{\infty}(r^l)|\le C r^l\le
%\frac{1}{2}r^k\le r^k$, and the desired  result follows from
%%
%$$ |r^k+O_{\infty}(r^l)|\ge|r^k-|O_{\infty}(r^l)||=r^k-|O_{\infty}(r^l)| \ge r^k-\frac{1}{2}r^k=\frac{1}{2}r^k\;.$$
%%
%\end{proof}
%%

 The Ernst equations~\eqref{Ernst} together with the estimates~\eqref{Ernst2},~\eqref{omega},~\eqref{dpsi} and~\eqref{psi},
that will be established in the next section, provide
\bel{Ernst2}
\left\{
 \begin{array}{l}
 \partial_z\newomega= -6J\rho^4/r^5+\rho\,O_{\infty}(\log r/r^3) \\
 \partial_{\rho}\newomega= 6J\rho^3z/r^5+\rho^2 O_{\infty}(\log r/r^4)
\end{array}\right.\;.
\ee
Integrating this system by using the polar coordinates
$\rho=r\cos\theta$, $z=r\sin\theta$, while imposing the standard
condition
\bel{initial cond}
\newomega(0,z)\equiv 0, \text{ for } z\gg0\;,
\ee
yields
 \bel{omega}
\newomega=4J-\frac{J}{2}\,\frac{z}{r}\left(\frac{3\rho^2-z^2}{r^2}+9\right)+\rho\, O_{\infty}(\log
r/r^2)\;.
 \ee

 We note the following relevant relation with the total angular
momentum as given by the Komar integral formula

\begin{align*}
\lim_{R\rightarrow\,+\infty}\frac{1}{16\pi} \int_{\{r=R\}}
*dK_\kl 1^{\flat}
          &=\lim_{R\rightarrow\,+\infty}-\frac{1}{16\pi}2\pi\int_{\{r=R\}\cap\{\varphi=0\}}i_{K_\kl 1}*dK_\kl 1^{\flat}  \\
          &=- \frac{1}{8}\lim_{R\rightarrow\,+\infty} \int_{\{r=R\}\cap\{\varphi=0\}} *(dK_\kl 1^{\flat}\wedge K_\kl 1^{\flat})\\
         % &=- \frac{1}{8}\lim_{R\rightarrow\,+\infty} \int_{\{r=R\}\cap\{\varphi=0\}} \omega\\
          &=- \frac{1}{8} \lim_{R\rightarrow\,+\infty} \int_{\{r=R\}\cap\{\varphi=0\}}\big( dv+2\underbrace{(\chi d\psi-\psi
d\chi)}_{=O(r^{-2})}\big) \\
          &=- \frac{1}{8} \lim_{R\rightarrow\,+\infty} \big(\newomega(0,R)-\newomega(0,-R)\big)\\
          &=- \frac{1}{8} \left(0-8J\right)\\
          &=J\;.
\end{align*}

We are now able to establish the electromagnetic twist potential part of~\eqref{goal2}.
For two twist potentials satisfying~\eqref{initial cond} we have
\bel{goal  solved1} \lim_{ \sqrt{\rho^2+z^2}\rightarrow +\infty \;,\rho\not\rightarrow 0} \frac{
\newomega_1-\newomega_2}{\rho^{2}}=0 \;. \ee

To take care of the asymptotic behavior of $\newomega$ near the axis
we Taylor expand on $\rho$ around a point $(0,z)$, away from the
poles, to get
 \bel{}
\newomega(\rho,z)=
\newomega(0,z)+\partial_{\rho}\newomega(c(\rho),z)\rho\;\;\;,\;\;
\abs{c(\rho)}\leq \rho\;.\ee
Then, using~\eqref{Ernst2} to obtain
 \bel{partial omega}
\partial_{\rho}\newomega= \rho^2O_{\infty}(r^{-3})\;,
\ee
we conclude that for $\abs{z}\gg0$ and $\rho\leq\abs{z}$
\bel{good for omega} \newomega=\newomega(0,z)+\rho^2O(r^{-3})\;. \ee
Finally, for two twist potentials that agree along the axis for both
large positive and negative $z$ we have, in the region
$\rho\leq\abs{z}$,
\bel{goal  solved2} \rho^{-2}(\newomega_1-\newomega_2)=O(r^{-3})\;,
\ee
and the desired result follows.

\subsubsection{The electromagnetic potentials}
\label{Selectro}

Asymptotic flatness~\eqref{A decay} together with~\eqref{A_0}
and~\eqref{tau} yield the desired improvement of the initial decay
rates
\bel{diA}
 \partial_{[i}A_{j]}=O_{\infty}(r^{-3})\;.
\ee
Now in the $\{x^{\mu}\}=\{t,x,y,z\}$ coordinates of the previous section we have
\begin{align*}
d\chi & := i_{K_\kl 1}F
\\
& =F_{\mu\nu}dx^{\mu}dx^{\nu}(K_\kl 1,\cdot)
\\
&= F_{\mu\nu}\left(dx^{\mu}(K_\kl 1)\:dx^{\nu}-dx^{\nu}(K_\kl
1)\:dx^{\mu}\right)=2 F_{\mu\nu}dx^{\mu}(K_\kl 1)\:dx^{\nu}
\\
&= 2 F_{\mu\nu}dx^{\mu}(x\partial_y-y\partial_x)\:dx^{\nu}=2
F_{\mu\nu}(x\delta^{\mu}_y-y\delta^{\mu}_x)\:dx^{\nu}
\\
&=2\left(xF_{y\nu}-yF_{x\nu}\right)\:dx^{\nu}=
4\left(x\partial_{[y}A_{\nu]}-y\partial_{[y}A_{\nu]}\right)\:dx^{\nu}
 \;.
\end{align*}
With~\eqref{diA} we see that, in the orbit space $\{\varphi=0\}$
(where $y=0$, $x=\rho$ and $\partial_x=\partial_{\rho}$), we have
 \bel{dchi}
 \left\{
 \begin{array}{l}
 \partial_{\rho}\chi|_{\varphi=0}=4\rho\, \partial_{[y}A_{\rho]}=\rho\, O_{\infty}(r^{-3}) \\
 \partial_{z}\chi|_{\varphi=0}=4\rho\, \partial_{[y}A_{z]}=\rho\, O_{\infty}(r^{-3})
\end{array}\right.\;.
\ee
Imposing the boundary condition
\bel{initial cond chi} \chi(0,z)\equiv 0, \text{ for } z\gg0\;, \ee
integration yields
\bel{chi control} \chi|_{\varphi=0}=\rho\,O_{\infty}(r^{-2})\;. \ee
\jlca{We're getting $\chi|_{\mcA}\equiv0$ without assuming the
existence of a global potential $A_{\mu}$  }

 Arguing as in the end of section~\ref{Stwist} the equation in~\eqref{goal2} corresponding to the potentials $\chi_i$ follows.

\bigskip

 To obtain a coordinate expression for $d\psi$ it will be helpful to rearrange
our preferred coordinate system and consider $\{x^{\mu}\}=\{t,y,x,z\}$, then
\begin{align*}
d\psi & := i_{K_\kl 1}*F
\\
&=\frac{1}{2}F^{\mu\nu}\epsilon_{\mu\nu\lambda\sigma}dx^{\lambda}dx^{\sigma}(K_\kl
1,\cdot) =
F^{\mu\nu}\epsilon_{\mu\nu\lambda\sigma}dx^{\lambda}(x\partial_y-y\partial_x)\:dx^{\sigma}
\\
&=F^{\mu\nu}\epsilon_{\mu\nu\lambda\sigma}(x\delta^{\lambda}_y-y\delta^{\lambda}_x)\:dx^{\sigma}
=F^{\mu\nu}(x\,\epsilon_{\mu\nu y\sigma}-y\,\epsilon_{\mu\nu
\,x\sigma})\:dx^{\sigma}\;.
\end{align*}

Now, in the orbit space and away from the axis, we have (compare
with~\eqref{space-time metric})
 \bel{gmunu}
 \fourg_{\mu\nu}|_{\varphi=0}=\left(
 \begin{array}{cccc}
 \fourg_{tt} & \rho^{-1}\fourg_{t \varphi} & 0 & 0 \\
 \rho^{-1}\fourg_{t \varphi} & \rho^{-2}\fourg_{\varphi \varphi} & 0 & 0 \\
 0 & 0 & e^{2u} & 0 \\
 0 & 0 & 0 & e^{2u}
\end{array}\right)\;,
\ee
therefore
\bel{det} \det(\fourg_{\mu\nu}|_{\varphi=0})=
\left(\fourg_{tt}\frac{\fourg_{\varphi
\varphi}}{\rho^{2}}-\frac{\fourg_{t
\varphi}^2}{\rho^{2}}\right)e^{4u}
=\frac{1}{\rho^2}(-\rho^2)e^{4u}=-e^{4u}\;; \ee
so
\begin{align*}
\partial_{\rho}\psi|_{\varphi=0} & = \rho\,\epsilon_{\mu \nu y x}F^{\mu\nu}=\rho(\epsilon_{t z y x}F^{tz}+\epsilon_{z t y x}F^{zt})
=2\rho\,\epsilon_{t z y x}F^{tz}
\\
&=2\rho\sqrt{|\det(\fourg_{\mu\nu})|}F^{tz}=2\rho\,
e^{2u}\fourg^{\mu t}\fourg^{\nu z}F_{\mu\nu}
\\
&=2\rho\, e^{2u}\fourg^{\mu t}\fourg^{z z}F_{\mu z}=2\rho\,
e^{2u}e^{-2u}(\fourg^{t t}F_{t z}+\fourg^{y t}F_{y z})
\\
&= 2\rho\left\{-\frac{\fourg_{\varphi \varphi}}{\rho^{2}}(\partial_t
A_z-\partial_zA_t)+2\frac{\fourg_{t
\varphi}}{\rho}\partial_{[y}A_{z]} \right\}
\\
&= 2\frac{\fourg_{\varphi \varphi}}{\rho}\partial_zA_t+4{\fourg_{t
\varphi}}\partial_{[y}A_{z]}
=2\rho(1+O_{\infty}(r^{-1}))\partial_zA_0+\rho \,O_{\infty}(r^{-5})
\;,
\end{align*}
where in the last equality we used~\eqref{estimate u},~\eqref{g phi
t} and~\eqref{diA}; also

\begin{align*}
\partial_{z}\psi|_{\varphi=0} & = \rho\,\epsilon_{\mu \nu y z}F^{\mu\nu}=2\rho\,\epsilon_{t x y z}F^{t x}
\\
&=-2\rho\sqrt{|\det(\fourg_{\mu\nu})|}F^{t x}=-2\rho\,
e^{2u}\fourg^{\mu t}\fourg^{\nu x}F_{\mu\nu}
\\
&=-2\rho\, e^{2u}\fourg^{\mu t}\fourg^{x x}F_{\mu x}=-2\rho\,
(\fourg^{t t}F_{t x}+\fourg^{y t}F_{y x})
\\
&= -2\rho\left\{-\frac{\fourg_{\varphi
\varphi}}{\rho^{2}}(\partial_t A_x-\partial_xA_t)+2\frac{\fourg_{t
\varphi}}{\rho}\partial_{[y}A_{x]} \right\}
\\
&=
-2\rho(1+O_{\infty}(r^{-1}))\partial_{\rho}A_0+\rho\,O_{\infty}(r^{-5})
\;.
\end{align*}
From~\eqref{A_0} we get
 \bel{dpsi}
 \left\{
 \begin{array}{l}
 \partial_{\rho}\psi|_{\varphi=0}=-Q\frac{\rho\,z}{r^3}+\rho\,O_{\infty}(r^{-3}) \\
 \partial_{z}\psi|_{\varphi=0}=Q\frac{\rho^2}{r^3}+\rho\,O_{\infty}(r^{-3})
\end{array}\right.\;.
\ee
Integrating as before while using a standard boundary condition
provides
 \bel{psi}
\psi=Q\left(-1+\frac{z}{r}\right)+\rho O_{\infty}(r^{-2})\;.
 \ee
We note the following relevant and expected relation with the {\em
total electric charge} given by the Komar integral
\bel{total charge}
 \lim_{R\rightarrow\,+\infty}-\frac{1}{4\pi} \int_{\{r=R\}} *F=Q\;.
\ee
%
%%
%\begin{align*}
% \lim_{R\rightarrow\,+\infty}-\frac{1}{4\pi} \int_{\{r=R\}} *F
%          &=\lim_{R\rightarrow\,+\infty}-\frac{-2\pi}{4\pi}\int_{\{r=R\}\cap\{\varphi=0\}}i_{K_\kl 1}*F  \\
%          &= \frac{1}{2}\lim_{R\rightarrow\,+\infty} \int_{\{r=R\}\cap\{\varphi=0\}} d\psi \\
%          &= \frac{1}{2} \lim_{R\rightarrow\,+\infty} \big(\psi(0,R)-\psi(0,-R)\big) \\
%          &= \frac{1}{2}Q \lim_{R\rightarrow\,+\infty} \big((-1+\frac{R}{R})-(-1+\frac{-R}{R})\big) \\
%          &=Q\;.
%\end{align*}
%%

It should be now clear that~\eqref{goal2} follows.

The results of this last two sections establish one of the
significant missing elements of all previous uniqueness claims for
the Kerr-Newman metric:

\begin{Proposition}
 \label{Tbdist}  Let $\Psi_i=(U_i, \newomega_i, \chi_i, \psi_i)$, $i=1,2$, be the Ernst
potentials associated with two $I^+$--regular, electro-vacuum,
stationary, asymptotically flat axisymmetric metrics with
non-degenerate event horizons. If $\newomega_1=\newomega_2$,
$\psi_1=\psi_2$ and $\chi_1=\chi_2$   on  the rotation axis, then
the hyperbolic-space distance between $\Psi_1$ and $\Psi_2$ is
bounded, going to zero as $r$ tends to infinity in the asymptotic
region.
\end{Proposition}

\section{Weinstein Solutions: existence and uniqueness}
 \label{sWeinstein}

 In this section we construct axisymmetric Ernst maps
$$\Phi=(U,\newomega,\chi,\psi):\R^3\setminus\mcA\rightarrow\HC\;, $$
which are ``close" to some reference maps, not necessarily harmonic, satisfying
conditions modeled on the local behavior of the Kerr-Newman
solutions. First recall the definitions of mass, angular momentum and electric charge of the $k$-th black hole as given by the Komar integrals
\bel{mass} m_k:=-\frac{1}{8\pi}\int_{S_k}*dK_\kl 0^{\flat}\;,\ee
\bel{angular momentum} J_k:=\frac{1}{16\pi}\int_{S_k}*dK_\kl
1^{\flat}\;,\ee
\bel{charge} q_k:=-\frac{1}{4\pi}\int_{S_k}*F\;.\ee
for some 2--sphere $S_k$ whose interior intersects the event horizon exactly at its $k$-th component.

We are now able to characterize the reference maps $\tilde\Phi=(\tilde U,\tilde\newomega,\tilde\chi,\tilde\psi)$:
\begin{enumerate}

\item
The components $\tilde f=\tilde \newomega, \tilde \chi$ and $\tilde \psi$ are locally bounded,
constant along each connected component of  $\mcA\setminus\mcE^+=\cup_{k=0}^N\mcA_k$ and we
write $\tilde f|_{\mcA_k}\equiv \tilde f_k$. These  functions are normalized to satisfy
$\tilde f_N= 0$.

\item
There exist $N_{\mbox{\scriptsize\rm  dh}}\ge 0$ degenerate event
horizons, which are represented by punctures $(\varphi=0,\rho=0,z=b_i)$,
together with a mass parameter $m_i>0$.
In a neighborhood of such  puncture, containing only this component of the horizon, the map $\tilde \Phi$ corresponds
to the harmonic map of the (extreme) Kerr-Newman solution parameterized by
$$(m_i,q_i)=(m_i,\frac{\tilde \psi_{i+1}-\tilde \psi_i}{2})\;.$$
%

%Near such punctures, i.e., for small
%$r_i:=\sqrt{\rho^2+(z-b_i)^2}$, the components of the map behave like
%%
%\begin{equation}
%  \label{eq:20e}
%U= \ln\Big(\frac{ r_i} {2m_i}\Big)+\frac 12 \ln\left( {1+
%\frac{(z-b_i)^2}{r_i^2} }\right)+ O(r_i)\;,
%\end{equation}
%%
%%
%\bel{}
%f= \mathring f(\theta) + o(1)\;,
%\ee
%%
%where $f=v,\chi,\psi$, and $\theta$ is the latitude of the spherical coordinates $(r_i,\varphi,\theta)$.

\item There exist $N_{\mbox{\scriptsize\rm  ndh}} \ge 0$
    non-degenerate horizons, which are represented by bounded open
    intervals $(c_i^-,c_i^+)=I_i\subset \mcA$, with none of the previous $b_j$'s
    belonging to the union of the closures of the $I_i$. In a neighborhood of such  interval, containing only this component of the horizon, the map $\tilde \Phi$ corresponds
to the harmonic map of the Kerr-Newman solution parameterized by
$$(\mu_j,\lambda_j,q_j)=(2\int_{I_j}dz,{\tilde v_{i+1}-\tilde v_i},\frac{\tilde \psi_{i+1}-\tilde \psi_i}{2})\;.$$
To retrieve the usual parametrization using mass, angular mommentum and charge one uses the known explicit formulas for Kerr-Newman
(e.g., equations 2.31. of~\cite{Weinstein96}) together with the following relations~\cite[section 2.3.]{Weinstein96}
\bel{parameters 1}
J_j=\frac{\lambda_j+l_j}{4}\;,
\ee
\bel{parameters 1}
m_j=\mu_j+2w_jJ_j\;,
\ee
where the auxiliary parameters are defined by $\lambda_j:=\int_{I_j}dv$, $l_j=\int_{I_j}\chi d\psi-\psi d\chi$, and $w|_{I_j}\equiv w_j$, with $w$ defined by~\eqref{space-time metric}.

    %The functions
%    $U-2\ln \rho$, $f=\newomega,\chi,\psi$ extend smoothly across each interval
%    $I_i$, with the following behavior near the end points, for some
%    constant $C$ (compare with~\eq{UbbhvOLD}):
%%
%\bel{hfin3} |U- \frac 12  \ln (\sqrt{\rho^2+(z-c_i^\pm)^2}+z-c_i^\pm
%)| \le C \;, \ee
% %
%%
%\bel{}
%f=O(1)\;,\quad
% \mbox{near $(0,0,c_i^\pm )$}\;.
%\ee
%%

%and jumps by $-4(J_k+\mu_k)$\jlca{4 should be 8!} when crossing the
%$k$-th component of $\cup _i\{b_i\}\cup_j I_j$ in the direction of
%increasing $z$, where
%%
%\bel{angular momentum} J_k:=\frac{1}{16\pi}\int_{S_k}*dK_\kl
%1^{\flat}\;\;\;\text{ and
%}\;\;\;\mu_k:=\frac{1}{16\pi}\int_{S_k}\chi d\psi-\psi d\chi\;,\ee
%%
%for some 2--sphere $S_k$ whose interior intersects $\cup
%_i\{b_i\}\cup_j I_j$ exactly at its $k$-th component.
%
%
%\item
%$\psi$ is locally bounded,
%constant along each connected component of $\mcA\setminus (\cup
%_i\{b_i\}\cup_j I_j)$, normalize to satisfy a condition analogous
%to~\eqref{initial cond}, and jumps by $2\,q_k$ when crossing the
%$k$-th component of $\cup_i\{b_i\}\cup_j I_j$ in the direction of
%increasing $z$, where $q_k$ is the Komar charge of the $k$-th
%component of the event horizon
%%
%\bel{charge} q_k:=-\frac{1}{4\pi}\int_{S_k}*F\;.\ee
%%
%
%\item
%$\chi$ is a locally bounded function which
%vanishes along $\mcA\setminus (\cup _i\{b_i\}\cup_j I_j)$.
%

\item In a neighborhood of infinity the functions $\tilde U$, $\tilde\newomega$, $\tilde\chi$ and $\tilde\psi$
coincide with the components of the harmonic map associated with the
Kerr-Newman solution with mass $M:=\sum_k m_k$ angular momentum
$J:=\sum_k J_k =v_0/8$ and electric charge $Q:=\sum_k q_k=-\psi_0/2$,  where the sums
are taken over all the components of the event horizon.

\item The functions $\tilde U$, $\tilde \newomega$, $\tilde \chi$ and $\tilde \psi$
are smooth across  $\mcA\setminus (\cup _i\{b_i\}\cup_j I_j)$.

\end{enumerate}

A collection $\{b_i,m_i\}_{i=1}^{\Ndh}$, $\{I_j, v(c^-_j),v(c^+_j)\}_{j=1}^{\Ndh}$,
and $\{\psi_k\}_{k=0}^{N-1}$
will be called \emph{``electro-vacuum axis data"}.

\bigskip

A map $\tilde\Phi$ satisfying condition 1.--5. above defines {\em
singular Dirichlet data}~\cite[Definition 2]{Weinstein:Hadamard} (compare~\cite[Section 2.4.]{Weinstein96})
with a target manifold with constant negative sectional curvature.
We then have the following version of~\cite[Theorem
2]{Weinstein:Hadamard} (compare~\cite[Appendix~C]{CLW} where the
uniqueness claim is clarified, and~\cite{Weinstein96} for a similar
result stated purely in terms of axis data):

\begin{Theorem}
 \label{Tuniquehm}
For any set of electro-vacuum axis data  there exists a unique
harmonic map $\Phi:\R^3\setminus \mcA\to \HC$ whose distance, as
given by~\eqref{dist}, from an axisymmetric map
$\tilde\Phi:\R^3\setminus \mcA\to \HC$, not necessarily harmonic but
with the properties 1.--5. above, satisfies:
\bel{unique criteria1} d(\Phi, \tilde\Phi)\in L^{\infty}(\R^3\setminus \mcA) %\text{ is bounded }
\;, \ee
and
\bel{unique criteria2} d(\Phi, \tilde\Phi)\rightarrow 0 \text{ as }
r\rightarrow +\infty \;. \ee
\qed
\end{Theorem}

%To construct the reference map $\Phi_0$ let $\{\mcU_i,
%\mcU_{\infty}\}$ be a cover of $(\cup _i\{b_i\}\cup_j
%I_j)\cup\{\infty\}$ by non-overlapping open sets, with the
%$i$-indices ordered in coherence with the respective $z$ axis
%position. Then, by shrinking the $\mcU_i$'s if necessary, we
%consider any map satisfying: $\Phi_0|_{\mcU_k}$ is the restriction
%to $\mcU_k$ of the harmonic map associated to the Kerr-Newman
%space-time with mass parameter $m_k$, angular momentum
%$\frac{v_{k+1}-v_{k}}{4}$\jlca{This needs checking!} and charge
%$\frac{\psi_{k+1}-\psi_{k}}{2}$;\jlca{How do we know that they
%satisfy the Kerr-Newman condition} $\Phi_0|_{\mcU_{\infty}}$ is the
%restriction to $\mcU_{\infty}$ of the harmonic map associated to the
%Kerr-Newman space-time satisfying 7.

 From an harmonic map $\Phi:\R^3\setminus \mcA\to \HC$ one can
construct a stationary and axisymmetric solution of the source free
Einstein-Maxwell field equations~\cite[Section 4.1.]{Weinstein96}.\jlca{with $\R\times(\R^3\setminus \mcA)$ topology. One than needs to extend across the axis, presumably in a singular way!!? How? what does this mean? Otherwise }
Such (not necessarily $I^+$--regular) space-times, arising from the
harmonic maps of the previous theorem will be referred to as {\em
Weinstein solutions}.

\section{Proof of Theorem~\ref{Tubh}}
\label{sproof}

 If $\mcEp$ is empty we obtain Minkowski by an aplication of~\cite[Theorem 2.7]{ChruscielNo}. Otherwise the proof splits into two
cases, according to whether or not $\changedX $ is tangent to the
generators of $\mcEp$.

\bigskip

{\bf Rotating horizons:}

Suppose, first, that the Killing vector  is not tangent to the
generators of some connected component $\mcE^+_0$. Proposition~1.9
of~\cite{ChruscielNo} allows us to generalize~\cite[Proposition
4.10]{ChrusCosta} to electro-vacuum and then Theorem~4.11 together
with the Remark~4.12 of~\cite{ChrusCosta} show that the event
horizon is analytic if the metric is; also, by~\eqref{diA} and
Einstein's equations, $G_{\mu\nu}=2\, T_{\mu\nu}=O(r^{-5})$. So the
Rigidity Theorem, as presented in~\cite[Theorem~5.1]{ChAscona},
applies and establishes the existence of a $\R\times \Uone$ subgroup
of the isometry group of $(\mcM,\fourg)$. The analysis of
Section~\ref{sWeyl}, leading to the global
representation~\eq{space-time metric} of the metric, is now
available. As stressed throughout this paper, in this gauge, the
field equations~\eqref{fieldeq1}-\eqref{fieldeq3} reduce to a
harmonic map $\Phi$~\eqref{harmonic map}. The analysis of the
asymptotic behavior of such map, whose results are compiled in
Proposition~\ref{Tbdist}, shows that $\Phi$ lies a finite distance
from one of the harmonic maps associated to the Weinstein solutions
of Theorem~\ref{Tuniquehm} and the uniqueness part of such theorem
allows us to conclude; note that in the connected and non-degenerate
setting the Weinstein solutions correspond to the non-extreme
Kerr-Newman metrics.

\bigskip

{\bf Non-rotating case:}

Now let us consider the case when the stationary Killing vector
$K_\kl 0$ is tangent to the generators of every component of
$\mcEp$. Following the procedure in~\cite[Section 7.2]{ChrusCosta},
based on~\cite{RaczWald2}, we extend $\doc$ to a space-time where
each connected component of the event horizon is contained in a
bifurcate horizon. Then, by~\cite{ChWald1} there exists an
asymptotically flat Cauchy hypersurface for the domain of outer
communications, with boundary on the union of the bifurcate spheres,
which is maximal. We are now able to conclude from Theorem~3.4
of~\cite{Sudarsky:wald} that $\doc$ is static. By taking in account
the corrections presented in the proof of Theorem~1.4
of~\cite{ChrusCosta} we can invoke, after relying on analyticity
once more, the non-degenerate part of the conclusion of Theorem~1.3
of~\cite{Chstaticelvac}, yielding non-extreme Reissner-Nordstr\"om
as the only non-rotating solution satisfying the remaining
conditions of the desired result.

\section{Concluding remarks}
 \label{sCr}

To obtain a satisfactory classification in four dimensions,
 the following issues remain to be addressed:

\begin{enumerate}
\item {\bf Analyticity.} The previous versions of the uniqueness theorem required
    analyticity of \emph{both} the metric \emph{and} the
    horizon. As shown in the proof of Theorem~\ref{Tubh}, the latter
    follows from the former. This is a worthwhile improvement,
    as even $C^1$-\emph{differentiability  of the horizon} is not clear a
    priori. But the hypothesis of analyticity of the metric remains to be removed.
In this context one should keep in mind the Curzon solution, where
analyticity of the metric fails precisely at the horizon.

 We further note that a new approach to Hawking's
rigidity without analyticity~\cite{IonescuKlainerman1,
IonescuKlainerman2} as yield significant breakthroughs in the vacuum
case.  According to {A.D.~Ionescu (private communication)} the
generalization of the results to electro-vacuum should follow by
similar techniques. However, some problems still need to be settled,
even for vacuum: the local claim requires a non-expanding horizon
which we expect to be a consequence of $I^+$--regularity and the
results and techniques of~\cite{ChrusCosta,ChDGH}, but such claim
requires checking; also, as it stands, the global result is
restricted to near Kerr geometries.

The hypothesis of analyticity is particularly annoying in the static
context, being needed there only to exclude non-embedded Killing
prehorizons~\cite[Section 5]{ChrusCosta}. The nature of that problem
seems to be rather different from Hawking's rigidity, with
presumably a simpler solution, yet to be found.

\item {\bf Degeneracy.} The classification of black holes with degenerate
    components of the event horizon requires further
    investigations. We believe that the
    results here go a long way to obtain a classification, in terms of Weinstein solutions, of
    stationary, axisymmetric, rotating
    configurations allowing both degenerate and non-degenerate components of the horizon:
    the foundations are settled but we are still missing an equivalent
    of Proposition~\ref{Tbdist}. Recall that in the static case
    a complete classification in terms of the
    Majumdar-Papapetrou and the Reissner-Norsdtr\"om families, with
    neither degeneracy or connectedness assumptions is already
    available by the work in~\cite{ChruscielTod} and references
    therein. In fact more is known in the degenerate class, since it
    was established in~\cite{CRT} that appropriately regular,
    $I^+$--regular in particular, Israel-Wilson-Perj\'es Black holes
    belong to the Majumdar-Papapetrou family.

 It has been
     announced~\cite{HorowitzDGT} that the question of
     uniqueness of degenerate black holes (with connected event horizon) has been settled.
     Unfortunately, that reference does not contain any new
     results, as compared to what had already been
     published in~\cite{ChrusCosta}, or is contained in this
     work, and so, it is our belief that this problem remains open. Indeed, the
     existence of global Weyl coordinates \emph{with
     controlled behavior at the singular set} is assumed.
     In the non-degenerate case this issue was first
     settled for vacuum in~\cite{ChrusCosta}, but the degenerate case appears
     to present serious technical difficulties, and
     requires further study.

\item {\bf Multi Component Solutions.}
    In agreement with the statement of
    Conjecture~\ref{Cubh}, one believes that all solutions with non-connected $\mcEp$
    are in the Majumdar-Papapetrou family. From what was said in the previous item, we see that
    it remains to show that non-static Weinstein solutions
    with non-connected horizons are singular;
    besides the already quoted result dealing with the Israel-Wilson-Perj\'es family,
    this has  been established  for
    slowly rotating black holes in vacuum\jlca{check the vacuum statement} by a regularity analysis
    of the relevant harmonic maps~\cite{LiTian,Weinstein:trans} and
    recent and promising results seem to have settled the problem for two-body configurations, also in vacuum~\cite{NH}.

\end{enumerate}

\bigskip

\noindent {\sc Acknowledgements:} We are grateful to Piotr Chru\'sciel and Jos\'e Nat\'ario for
numerous comments on a previous version of the paper and many useful discussions.

%%%%%%%%%%%%%%%%%%%%%%%%%%%%%%%%%%%%%%%%%%%%%%%%%%%%%%%%%%%%%%%%%%%%%%%%%%%%%%%%%%%%%%%%%%%%%%%%%%%%%%%%%%%%%%%%%%
%%%%%%%%%%%%%%%%%%%%%%%%%%%%%%%%%%%%%%%%%%%%%%%%%%%%%%%%%%%%%%%%%%%%%%%%%%%%%%%%%%%%%%%%%%%%%%%%%%%%%%%%%%%%%%%%%%
%%%%%%%%%%%%%%%%%%%%%%%%%%%%%%%%%%%%%     BIBLIO     %%%%%%%%%%%%%%%%%%%%%%%%%%%%%%%%%%%%%%%%%%%%%%%%%%%%%%%%%%%%%
%%%%%%%%%%%%%%%%%%%%%%%%%%%%%%%%%%%%%%%%%%%%%%%%%%%%%%%%%%%%%%%%%%%%%%%%%%%%%%%%%%%%%%%%%%%%%%%%%%%%%%%%%%%%%%%%%%

\providecommand{\bysame}{\leavevmode ---\ } \providecommand{\og}{``}
\providecommand{\fg}{''} \providecommand{\smfandname}{et}
\providecommand{\smfedsname}{\'eds.}

 \end{document}